 \documentclass[final,3p,times,twocolumn]{elsarticle}

 \usepackage{graphicx}

\usepackage{amssymb}
\usepackage{amsmath}

\journal{Physica A}

\begin{document}

\begin{frontmatter}



\title{The relativistic kinetic dispersion relation:
Comparison of the relativistic Bhatnagar-Gross-Krook model and Grad's 14-moment expansion}


\author{Makoto Takamoto}

\address{Department of Physics, Kyoto University, Kyoto, 606-8502, Japan}

\author{Shu-ichiro Inutsuka}

\address{Department of Physics, Nagoya University, Nagoya, 464-8602, Japan}

\begin{abstract}
In this paper, we study the Cauchy problem of the linearized kinetic equations 
for the models of Marle and Anderson-Witting, 
and compare these dispersion relations with the 14-moment theory.
First, we propose a modification of the Marle model 
to improve the resultant transport coefficients in accord with 
those obtained by the full Boltzmann equation. 
Using the modified Marle model and Anderson-Witting model, 
we calculate dispersion relations that are kinetically correct within the validity of the BGK approximation. 
The 14-moment theory that includes the time derivative of dissipation currents 
has causal structure, 
in contrast to the acausal first-order Chapman-Enskog approximation. 
However, the dispersion relation of the 14-moment theory does not accurately describe 
the result of the kinetic equation. 
Thus, 
our calculation indicates that 
keeping these second-order terms does not simply 
correspond to improving the physical description of the relativistic hydrodynamics. 
\end{abstract}


\end{frontmatter}


\section{\label{sec:level0}INTRODUCTION}
Recently, the interest in relativistic dissipative fluids in astrophysics and nuclear physics 
has increased. 
Relativistic dissipative fluid equations have many features that do not appear in the case of nonrelativistic fluid.
The most basic difference is the fact that in the presence of heat flux, 
the fluid velocity cannot be defined uniquely. 
There are two well-known definitions of fluid velocity; 
Eckart velocity~\cite{Eckart(1940)} that is parallel to particle flow
and Landau-Lifshitz velocity~\cite{Landau & Lifshitz(1959)} that is parallel to energy flow.
In addition, 
it is well known that standard first-order relativistic Navier-Stokes hydrodynamics exhibits fatal problems regarding causality and stability, 
that is, small perturbations to the uniform static states grow exponentially~\cite{Hiscock & Lindblom(1983),Hiscock & Lindblom(1985)}. 
Currently, 
the most widely accepted and studied theory is the second-order Israel-Stewart (IS) approach~\cite{Israel & Stewart(1979)}
based on the 14-moment method~\cite{Cerbook}.
Unfortunately, 
this theory is inconvenient for practical use 
because we have to restore so many terms that are second-order in deviations from equilibrium, 
that is, the time derivative of dissipation terms 
and the products of gradients of dissipative quantities. 
However, because of the recent finding of the strongly coupled 
quark-gluon plasma (sQGP) in the Relativistic Heavy-Ion Collider (RHIC), 
description by relativistic hydrodynamics equations have been vigorously studied in the context of nuclear physics~\cite{Baier et al.(2008)}, 
and application of IS theory has just begun~\cite{Molnar et al.(2009)}.
Recently, a new approach to relativistic dissipative fluid equation has been shown by Tsumura, Kunihiro, and Ohnishi~
\cite{Tsumura et al.(2007),Tsumura & Kunihiro(2008),Tsumura & Kunihiro(2009)}.
They use the renormalization-group method for obtaining fluid equation from the Boltzmann equation, 
and obtained equation is different from both Eckart and Landau-Lifshitz equation.

Microscopic phenomena are accurately described by the Boltzmann equation.
However, it is very difficult to solve 
since its collision term depends on the product of the distribution functions.
Consequently, a simpler approximation for the collision term has been proposed; 
the most widely used relativistic kinetic model equations are 
those of Marle~\cite{Marle} and Anderson-Witting~\cite{Anderson & Witting(1974)}.
The Marle model is an extension of the nonrelativistic 
Bhatnagar-Gross-Krook (BGK) model~\cite{Bhatnagar et al.(1954)} to the relativistic case 
and is described in the Eckart frame~\cite{Eckart(1940)}. 
The Anderson-Witting model is another extension 
and is described in the Landau-Lifshitz frame~\cite{Landau & Lifshitz(1959)}.
Of the two, the Anderson-Witting model is widely used~\cite{Struchtrup(1998),Yano et al.(2007)}
because the Marle model has undesirable properties; for example, 
the transport coefficients obtained by the Marle model 
do not agree with those obtained by the full Boltzmann equation~\cite{Cerbook}.

In this paper,
we compare the dynamics described by the 14-moment theory with that of the kinetic model equation 
and test the applicability of the IS approach. 
To do numerical simulation of relativistic dissipative fluid, 
we should know how to treat the small second-order terms 
and how to determine appropriate values of new coefficients, 
which urges us to check how important these terms are.
To make the problem tractable, 
we study linear perturbation and compare the solutions of the dispersion relation.
The dispersion relations of the relativistic kinetic equations have been studied as a boundary value problem 
by Cercignani and Majorana~\cite{cer1984,Cercignani & Majorana(1985)}. 
To understand the dynamics as a Cauchy problem, 
we solve the dispersion relations with respect to $\omega$.
In addition, 
we modify the problematic properties of the Marle model 
and use the modified model equation to analyze the Eckart description.

%
This paper is organized as follows: 
in Sec.~\ref{sec:level1}, 
we introduce the kinetic models of Marle and Anderson-Witting. 
Then, we modify the Marle model 
and obtain the dispersion relations.
In Sec.~\ref{sec:level2}, 
we solve the dispersion relations numerically with respect to $\omega$ 
and present our results. 
In Sec.~\ref{sec:level3}, 
we discuss the properties of the Marle and Anderson-Witting models. 
In addition, 
we analyze the asymptotic behavior of the dispersion relations. 
First, we study the long wavelength limit 
and then we study the short wavelength and high frequency limits.
We solve the dispersion relation of the 14-moment theory 
and compare it with the dispersion relations of the kinetic model equation.


\section{\label{sec:level1}THE LINEARIZED KINETIC EQUATIONS AND THE DISPERSION RELATIONS}
In this section, we derive the dispersion relations of the relativistic kinetic models of Marle and Anderson-Witting.
Throughout this paper, we use the units
\begin{equation}
c = 1, \quad k_B = 1
,
\end{equation}
where $c$ is the velocity of light, and $k_B$ the Boltzmann constant.

In Cartesian coordinates, the Minkowski metric tensor $\eta_{\mu \nu}$ is given by
\begin{equation}
\eta_{\mu \nu} = \mathrm{diag}(1, -1, -1, -1)
.
\end{equation}
Variables indicated by Greek letters take values from $0$ to $3$, 
and those indicated by Roman letters take values from $1$ to $3$.

\subsection{\label{sec:mlevel0}MODIFICATION OF THE BGK MODEL OF MARLE}
Marle~\cite{Marle} has proposed the following form of the kinetic model equation,
\begin{equation}
\left(\frac{\partial f}{\partial t} \right)_{coll} = 
- \frac{m}{\tau_M} (f(t,{\bf x},{\bf p})-f_{eq}(t,{\bf x},{\bf p})) 
,
\label{eq:mBGK}
\end{equation}
where $\tau_M$ is a characteristic time on the order of the mean flight time 
(see below for its physical interpretation), 
$m$ is the rest mass of a particle of the relativistic gas, and 
$f_{eq}$ is the local equilibrium distribution function.

Using Eq.~(\ref{eq:mBGK}), 
we obtain the following form of the kinetic equation
\begin{align}
p^{\mu} \partial_{\mu} f &= p^0 \left(\frac{\partial}{\partial t} + {\bf v} \cdot \nabla \right) f 
= - \frac{m}{\tau_M} (f-f_{eq}) 
,
\label{eq:mBGKbol} \\
{\bf v} &= \frac{{\bf p}}{p^0}
.
\end{align}

The Marle model is an extension of the nonrelativistic BGK model to the relativistic case.
The transport coefficients for the Marle model equation reproduce the nonrelativistic results 
in the limiting case of low temperature.
It is, however, well known that in the limiting case of high temperature, 
the transport coefficients of the Marle model
differ from those found for hard-sphere particles obtained by the full Boltzmann equation~\cite{Cerbook}.
More precisely, if we express the transport coefficients ($\propto \tau_M$) as a function of $\zeta = m / T$, 
the transport coefficients of the Marle model behave as $1 / \zeta$ of those found 
by the Boltzmann equation for hard-sphere particles in the limit of high temperature.
For this problem, we should recall that the transport coefficients are generally proportional to the relaxation time $\tau_M$, 
and Eq.~(\ref{eq:mBGK}) contains $\tau_M$ as a parameter of the BGK model.
This indicates that 
the appropriate value of $\tau_M$ is different from the physical relaxation timescale $\tau_{relax}$ by a factor 
that becomes unity in the low temperature 
limit and becomes $\zeta$ in the high temperature limit. 
We discuss this new interpretation of $\tau_M$.

First, 
we clarify the meaning of the parameter $\tau$ in the BGK model.
In the nonrelativistic BGK model, 
the parameter $\tau$ is equivalent to the relaxation time.
The nonrelativistic kinetic equation of the BGK model is
\begin{equation}
\left(\frac{\partial}{\partial t} + {\bf v} \cdot \nabla \right) f 
= - \frac{1}{\tau} (f(t,{\bf x},{\bf v})-f_{eq}(t,{\bf x},{\bf v})) 
.
\label{eq:BGKbol}
\end{equation}
If the one-particle distribution function $f$ does not depend on the spatial coordinates, 
Eq.~(\ref{eq:BGKbol}) reduces to the ordinary first-order differential equation, 
and we can obtain the formal solution 
\begin{equation}
f(t) = \left[f(0) + \frac{1}{\tau} \int^t_0 e^{t' / \tau} f_{eq}(t') dt' \right] e^{-t/\tau}
.
\end{equation}
This equation indicates that $\tau$ is the relaxation time of the distribution function.

Next, 
we consider the relativistic BGK model of Marle. 
The kinetic equation of the Marle model is
\begin{align}
&p^{\mu} \partial_{\mu} f = p^0 \left(\frac{\partial}{\partial t} + {\bf v} \cdot \nabla \right) f
\nonumber
\\
&= - \frac{m}{\tau_M} \left(f(t,{\bf x, v}) - f_{eq}(t, {\bf x, v}) \right)
.
\label{eq:mBGK2}
\end{align}

If we assume that the one-particle distribution function $f$ does not depend on the spatial coordinates, 
the formal solution of Eq.~(\ref{eq:mBGK2}) is
\begin{align}
f(t) &= \left[f(0) + \frac{1}{\tau_{M*}} \int^t_0 e^{t'/\tau_{M*}} f_{eq}(t') dt' \right] e^{-t/\tau_{M*}}
\label{eq:Mrelax}
,
\\
\tau_{M*} &= \frac{p^0}{m} \tau_M
.
\end{align}
This indicates that in a general inertial frame, 
the relaxation time is not $\tau_M$ but $\tau_{M*}$, and 
$\tau_M$ is the relaxation time in the rest frame where the momentum of particles is ${\bf p = 0}$.
More precisely, 
if we employ the particle's rest frame where {\bf p = 0}, 
Eq.~(\ref{eq:mBGK2}) becomes
\begin{equation}
\frac{\partial}{\partial t} f(t,{\bf x,0}) = - \frac{m}{\tau_M} (f(t,{\bf x},{\bf 0})-f_{eq}(t,{\bf x},{\bf 0}))
.
\end{equation}
This is the same equation as in the nonrelativistic BGK model, 
indicating that only in this frame does $\tau_M$ become the relaxation time.

Although the transport coefficients of the Marle model are expressed in a form proportional to $\tau_M$ in the literature~\cite{Cerbook},
the above explanation shows that we should use $\tau_{M*}$ as the relaxation time instead of $\tau_M$.
However, 
$\tau_{M*}$ depends on the momentum $p^0$, so $\tau_{M*}$ cannot appear in macroscopic descriptions, 
such as transport coefficients.
For this reason, we have to consider the true relaxation time $\tau_{relax}$, 
to which the transport coefficients should be proportional, and
relate it to the BGK parameter of the Marle model $\tau_M$.
The above discussion suggests that we may regard $1 / \tau_{relax}$ as $\left\langle 1 / \tau_{M*} \right\rangle$, 
and we can consider $\tau_{relax}$ as the effective relaxation time in general frames. 
%
%
Using the local equilibrium distribution function, 
$\tau_M$ is
\begin{equation}
\tau_M = \frac{m}{n} \int \frac{d^3 p}{p^0} f_{eq} \tau_{relax} = \frac{K_1(\zeta)}{K_2(\zeta)} \tau_{relax} 
\label{eq:taum}
,
\end{equation}
where $K_n$ is the second kind modified Bessel function of order n.
The correction $K_1(\zeta)/K_2(\zeta)$ becomes $1$ in the limit of large $\zeta$ and 
$\zeta/ 2$ when $\zeta$ is nearly $0$.
This indicates that this function has the desired properties.
In the following, we use this $\tau_M$ as the BGK parameter of the Marle model.

In above discussion, 
we assume that the physical system is not far from equilibrium state, and 
calculate the average of $\tau_M$ with respect to the local equilibrium distribution function $f_{eq}$.
Though this cannot give the correct $\tau_M$ in the general case,
it is a good approximation for linear perturbation about the local equilibrium distribution function.


\subsection{\label{sec:mlevel1}THE LINEARIZED KINETIC EQUATION AND 
DISPERSION RELATION OF THE MODIFIED MARLE MODEL}
In this section, 
we derive the dispersion relation of the  modified relativistic kinetic model of Marle.
To obtain the dispersion relation, 
we apply an approach similar to that in the work of Cercignani and Majorana~\cite{cer1984}.

When there is no external field, the equation of the modified Marle model is given by
\begin{eqnarray}
&& \frac{D}{D s} f = - \frac{m}{\tau_M} (f-f_{eq}) 
,
\label{eq:mBGKbolkai}
\\
&& \frac{D}{D s} f = p^{\mu} \partial_{\mu} f = p^0 \left(\frac{\partial}{\partial t} + {\bf v} \cdot \nabla \right) f
.
\end{eqnarray}

In Eq.~(\ref{eq:mBGKbolkai}), $\tau_M$ is the relaxation time modified in Sec.~\ref{sec:mlevel0}, and 
$f_{eq}$ represents the local Maxwell-J$\ddot{\mathrm{u}}$ttner distribution function
\begin{eqnarray}
f_{eq}(t,{\bf x, p}) &=& \frac{n(t,{\bf x})}{4 \pi m^2 T(t,{\bf x}) K_2(\zeta(t,{\bf x}))} 
\nonumber
\\
&\times&\exp \left[- \frac{p_{\mu} u^{\mu}(t,{\bf x})} {T(t,{\bf x})}\right] 
\label{eq:mjeq}
,
\\
\zeta &=& \frac{m}{T}
,
\end{eqnarray}
where $m$ is the mass of the particle, and $T$ is the temperature.

Eq.~(\ref{eq:mBGKbolkai}) is a nonlinear equation for $f(t,{\bf x, p})$ because of 
the nonlinear dependence of $f_{eq}$ on $f$ through the following conditions called the matching conditions:
\begin{eqnarray}
&&\int \left( f_{eq} - f \right) \psi \frac{d^3 p}{p^0} = 0  
\label{eq:mrmat}
,
\\
&&\psi = (1, ~p^{\mu})
\label{eq:psi}
.
\end{eqnarray}

To obtain the dispersion relation, 
we start by expanding the distribution function around a global equilibrium state $f_0({\bf p})$, 
\begin{eqnarray}
\delta f = f- f_0, \quad \delta f_{eq} = f_{eq} - f_0.
\end{eqnarray}
Then, the linearized kinetic equation of the modified Marle model is given by
\begin{eqnarray}
\left( \frac{\partial}{\partial t} + {\bf v} \cdot \nabla \right) \delta f  
&=& - \frac{\delta f - \delta f_{eq}}{\tau_{M*}}
,
 \label{eq:mlinbol} \\
\tau_{M*} &=& \frac{p^0}{m} \tau_M
.
\end{eqnarray}

We assume a solution in the following form:
\begin{equation}
\delta f = \delta \tilde{f} e^{- i k_{\mu} x^{\mu}} = \delta \tilde{f} e^{- i \omega(t-t_0)+i {\bf k \cdot x}} 
.
\label{eq:mome}
\end{equation}
Then, Eq.~(\ref{eq:mlinbol}) reduces to
\begin{equation}
\left( \frac{1}{\tau_{M*}} -i \omega + i {\bf k \cdot v} \right) \delta f = \frac{1}{\tau_{M*}} \delta f_{eq}
\label{eq:mlinbolf}
.
\end{equation}
We consider an equilibrium background state in which the fluid is at rest, so that $u^{\mu} = (1, {\bf 0})$ and 
$\delta u^{\mu} = (0, \delta {\bf u})$ owing to the relation $u^{\mu} \delta u_{\mu} = 0$.
Then, $\delta f_{eq}$ is given by
\begin{align}
\delta f_{eq} &= f_0 
\nonumber
\\
&\times \left[ \frac{\delta n}{n} + \left(-1 + \frac{p^0}{T} + \frac{K_2'}{K_2} \zeta \right) \frac{\delta T}{T}
 - \frac{{\bf p} \cdot \delta {\bf u}}{T} \right] 
,
\\
f_0 &= \frac{n}{4 \pi m^2 T K_2(\zeta)} \exp \left[- \frac{p^0}{T} \right]
,
\end{align}
where $\delta {\bf u}$ is the space component of the Eckart velocity, as explained in Sec.~\ref{sec:level31}, and 
$K_n'$ is the derivative of $K_n$ with respect to $\zeta$.

Using the matching conditions, we can rewrite $\delta \rho, \delta {\bf u}$, and $\delta T$ as the integrals of $\delta f$:
\begin{align}
\delta n(t,{\bf x}) &= \int  \frac{d^3 p}{p^0} p^0 \delta f
, \\
\delta {\bf u}(t,{\bf x}) &= - \frac{1}{n} \int  \frac{d^3 p}{p^0} {\bf p} \delta f
, \\
\delta T(t,{\bf x}) &= \int \frac{d^3 p}{p^0} p^0 
\\
\nonumber
&\times
\frac{- 1 + K_2'  \zeta / K_2 + p^0 / T}{\left(1 - K_1  \zeta / K_2 \right)
\left(3 + \zeta^2 + K_1 \zeta / K_2 \right)} \delta f
.
\end{align}

Eq.~(\ref{eq:mlinbolf}) becomes
\begin{align}
& \left( \frac{1}{\tau_{M*}} -i \omega + i {\bf k \cdot v} \right) \delta f({\bf p}) 
\\ \nonumber
&= \int \frac{d^3 p'}{p'^0} \frac{f_0({\bf p})}{\tau_{M*}} 
\left[\frac{p'^0}{n} - \frac{{\bf p} \cdot {\bf p'}}{T}
\right.
\\ \nonumber
&+ \left.
 \frac{T}{n} \frac{p'^0}{\left(1-K_1 \zeta / K_2\right) \left(3 + \zeta^2 + K_1 \zeta / K_2 \right)} 
\right.
\\ \nonumber
&\times \left. \left(-1 + \frac{p^0}{T} + \frac{K_2'}{K_2} \zeta \right) 
 \left( - 1 + \frac{p'^0}{T} + \frac{K_2'}{K_2} \zeta \right)  \right] \delta f({\bf p'})
.
\end{align}

In the following, 
we take $\tau_{relax}$ as a unit of time: 
\begin{equation}
\omega \tau_{relax} \rightarrow \omega, \quad \tau_{relax} k \rightarrow k.
\end{equation}

Finally, the linearized equation of the BGK model of Marle is
\begin{align}
\delta f({\bf p}) &= \int \frac{d^3 p'}{p'^0} K({\bf p,p'}) \delta f({\bf p'}) 
\label{eq:mlinbolk}
,
\\
K({\bf p,p'}) &\equiv \frac{f_0({\bf p})}{1 - \left(i \omega - i {\bf k} \cdot \frac{{\bf p}}{p^0} \right)
 \frac{K_1 z}{K_2 \zeta}}
\left[\frac{p'^0}{n} - \frac{{\bf p} \cdot {\bf p'}}{T}
\right.
\\ \nonumber
&+ \left. \frac{T}{n} \frac{p'^0}{\left(1-K_1 \zeta / K_2\right) \left(3 + \zeta^2 + K_1 \zeta / K_2 \right)} 
\right.
\\ \nonumber
&\times \left. \left(-1 + z + \frac{K_2'}{K_2} \zeta \right) 
 \left( - 1 + z' + \frac{K_2'}{K_2} \zeta \right)  \right] \delta f({\bf p'})
,
\end{align}
where $z = p^0 / T$. 
This equation make sense only when 
$1 - \left(i \omega - i {\bf k} \cdot \frac{{\bf p}}{p^0} \right) \frac{K_1 z}{K_2 \zeta} \neq 0$; 
we explain the case where 
$1 - \left(i \omega - i {\bf k} \cdot \frac{{\bf p}}{p^0} \right) \frac{K_1 z}{K_2 \zeta} = 0$ 
later.

Eq.~(\ref{eq:mlinbolk}) is the homogeneous Fredholm integral equation of the second kind.
In particular, the kernel function $K({\bf p, p'})$ can be separated with respect to the variables ${\bf p}$ and ${\bf p'}$. 
Thus, this equation can be solved according to a general procedure.

First, we integrate Eq.~(\ref{eq:mlinbolk}) with respect to ${\bf p}$.
Then, we multiply by $K_1 k / \zeta$
and the equation reduces to
\begin{equation}
I_{11} \frac{\delta n}{n}  + I_{12} {\bf k} \cdot \delta {\bf u} + I_{13} \frac{\delta T}{T} = 0
,
\label{eq:mcont}
\end{equation}
where
\begin{align}
I_{11} &= K_1^2 k - \zeta K_2 P(0) - \pi K_2 e^{-d}
,
\\
I_{12} &= -\frac{i K_2 \zeta}{k} \left\{\frac{\zeta}{k} \left(\frac{K_2}{K_1} P(0) - i \omega P(1) \right) - K_1 \right\}
\\ \nonumber
&- \pi K_2 e^{-d} \frac{i}{k^2} \left\{-i \omega (1 + d) + \frac{K_2}{K_1} \zeta \right\}
,
\\
I_{13} &= \left(-3-\frac{K_1}{K_2} \zeta \right) \left(k K_1^2 - \zeta K_2 P(0) \right) 
\\ \nonumber
&+ 
\zeta K_2 \left(k K_1 - \zeta P(1) \right) 
- \pi K_2 e^{-d} \left(d - 2 - \frac{K_1}{K_2} \zeta \right)
,
\end{align}
$P(n)$ is
\begin{align}
P(n) &= \int^{\infty}_{1} dy \; e^{-\zeta y} y^n \arctan \frac{\zeta \sqrt{y^2 - 1}}{b}
,
 \\
b &= \frac{\zeta}{k} \left( \frac{K_2}{K_1} - i \omega y \right)
,
\end{align}
and $d$ is
\begin{align}
d &= - \frac{K_2 \zeta}{K_1 \mathrm{Im}(\omega)} \qquad \mathrm{if} \quad  \mathrm{Im}(\omega) < - \frac{K_2}{K_1}
,
\\ \nonumber
d &= \zeta \qquad \qquad \qquad \mathrm{if} \quad \mathrm{Im}(\omega) > - \frac{K_2}{K_1} 
.
\end{align}
The derivation of the correction term $c$ is explained in Sec.~\ref{sec:mlevela2}.

Next, we multiply Eq.~(\ref{eq:mlinbolk}) by ${\bf k \cdot p} \equiv k p^x$ and 
integrate with respect to ${\bf p}$.
Then, we multiply by $\zeta K_1$ 
and the equation reduces to
\begin{equation}
I_{21} \frac{\delta n}{n} + I_{22} {\bf k} \cdot {\bf \delta u} + I_{23} \frac{\delta T}{T} = 0
\label{eq:mmom}
,
\end{equation}
where
\begin{align}
I_{21} &= \zeta \left(- i K_1 + i \frac{K_2 \zeta}{K_1 k} P(0) + \frac{\omega \zeta}{k} P(1) \right) 
\\ \nonumber
&+ \frac{\pi}{k} e^{-d} 
\left\{ \omega (1 + d) + i \frac{K_2}{K_1} \zeta \right\}
,
\\
I_{22} &= \frac{\zeta^2}{k^2} \left\{K_2 (1 - i \omega) - \frac{K_2 \zeta}{K_1 k} \left(\frac{K_2}{K_1} P(0) - i \omega P(1) \right) \right.
\\ \nonumber
&+ \left. \frac{i \omega \zeta}{k} \left(\frac{K_2}{K_1} P(1) - i \omega P(2) \right) - \zeta K_1 \right\}
\\ \nonumber
&+ \frac{\pi}{k} e^{-d} \frac{1}{k^2} \left\{\omega^2 (d^2 + 2 d + 2) 
\right.
\\ \nonumber
&+ \left. 
2 i \omega \frac{K_2}{K_1} \zeta (1 + d) - 
\left(\frac{K_2 \zeta}{K_1} \right)^2 \right\}
,
\\
I_{23} &= - i \zeta \left(- 3 K_1 - \frac{K_1^2}{K_2} \zeta + \zeta K_2 \right) 
\\ \nonumber
&+ i \frac{K_2 \zeta^2}{K_1 k} \left(- 3 P(0) - \frac{K_1 \zeta}
{K_2} P(0) + \zeta P(1) \right)
\\ \nonumber
&+ \frac{\omega \zeta^2}{k} \left(- 3 P(1) - \frac{K_1 \zeta}{K_2} P(1) + \zeta P(2) \right)
\\ \nonumber
&+  \frac{\pi}{k} e^{-d} \left[\omega \left\{(1 + d) \frac{K_2'}{K_2} \zeta + d^2 + d + 1 \right\} 
\right.
\\ \nonumber
&+ \left. i \frac{K_2}{K_1} \zeta \left(
d + \frac{K_2'}{K_2} \zeta \right) \right]
. 
\end{align}

Next, we multiply Eq.~(\ref{eq:mlinbolk}) by ${\bf k \times p} \equiv k p^{\perp}$ and 
integrate with respect to ${\bf p}$.
Then we multiply by $K_1 k$
and the equation reduces to
\begin{equation}
I_{\perp \perp} \delta u_{\perp} = 0
,
\label{eq:mshear} 
\end{equation}
where
\begin{align}
I_{\perp \perp} &= 2 k K_1 - \int^{\infty}_1 dy \; e^{-\zeta y} \left[ -b \zeta \sqrt{y^2 - 1} 
\right.
\\ 
\nonumber
&+ \left. \left\{ b^2 + \zeta^2 (y^2 - 1) \right\} 
\arctan \frac{\zeta \sqrt{y^2 - 1}}{b} \right]
\\ 
\nonumber
&- \frac{\pi}{\zeta} e^{-d} \left\{(d^2 + 2 d + 2) \left(1 - \frac{\omega^2}{k^2} \right) 
\right.
\\
\nonumber
&- \left. \frac{2 i K_2}{k^2 K_1} \zeta \omega (1 + d)
+ \left(\frac{K_2^2 \zeta^2}{k^2 K_1^2} - \zeta^2 \right) \right\}
,
\end{align}
Eq.~(\ref{eq:mmom}) corresponds to the longitudinal mode ($\delta u_x \neq 0, \delta u_{\perp} = 0$), 
and Eq.~(\ref{eq:mshear}) corresponds to the transverse mode ($\delta u_x = 0, \delta u_{\perp} \neq 0$).

Finally, we multiply Eq.~(\ref{eq:mlinbolk}) by $p^0$ and 
integrate with respect to ${\bf p}$.
We multiply by $K_1 k$ and 
the equation reduces to
\begin{equation}
I_{31} \frac{\delta n}{n} + I_{32} {\bf k} \cdot \delta {\bf u} + I_{33} \frac{\delta T}{T} = 0
,
\label{eq:men}
\end{equation}
where
\begin{align}
I_{31} &= \zeta P(1) - K_1 k + \frac{\pi}{\zeta} e^{-d} (1 + d)
,
\\
I_{32} &= - \frac{i \zeta}{k} \left\{K_2 - \frac{\zeta}{k} \left(\frac{K_2}{K_1} P(1) - i \omega P(2) \right) \right\}
\\
\nonumber
&+ \frac{\pi}{\zeta} e^{-d} \frac{i}{k} \left\{- i \omega (d^2 + 2 d + 2) + \frac{K_2}{K_1} \zeta (1 + d) \right\}
,
\\
I_{33} &= \zeta \left(- 3 P(1) - \frac{K_1}{K_2} \zeta P(1) + \zeta P(2) \right) 
\\
\nonumber
&+ \frac{\pi}{\zeta} e^{-d} \left\{(1 + d) \frac{K_2'}{K_2} \zeta + d^2 + d + 1 \right\}
.
\end{align}

If the determinant of the above homogeneous system is set to equal zero,
the following dispersion relation is obtained:
\begin{eqnarray}
 \begin{vmatrix}
   I_{11} & I_{12} & I_{13} & 0 & 0 
   \\
   I_{21} & I_{22} & I_{23} & 0 & 0 
   \\
   I_{31} & I_{32} & I_{33} & 0 & 0
   \\
   0 & 0 & 0 & I_{\perp \perp} & 0 
   \\
   0 & 0 & 0 & 0 & I_{\perp \perp} 
 \end{vmatrix}
= 0
.
\label{eq:det1}
\end{eqnarray}
This condition implies either
\begin{eqnarray}
 \begin{vmatrix}
   I_{11} & I_{12} & I_{13}
   \\
   I_{21} & I_{22} & I_{23}
   \\
   I_{31} & I_{32} & I_{33}
 \end{vmatrix}
= 0
,  
\label{eq:det2}
\end{eqnarray}
or
\begin{equation}
I_{\perp \perp} =0
.
\label{eq:det3}
\end{equation}

Using this dispersion relation, 
we can obtain $\delta f$ in the form 
\begin{align}
\delta f({\bf v}) &= \sum_n \frac{C_n f_0({\bf v})}{1 - \left(i \omega - i {\bf k} \cdot \frac{{\bf p}}{p^0} \right)
 \frac{K_1 z}{K_2 \zeta}} 
\label{eq:mgenesol}
\\
\nonumber
&\times
\left[ \frac{\delta n_{\omega_n}}{n} + \left(-1 + \frac{p^0}{T} + \frac{K_2'}{K_2} \zeta \right) \frac{\delta T_{\omega_n}}{T}
\right.
\\
\nonumber
&- \left. \frac{{\bf p} \cdot \delta {\bf u}_{\omega_n}}{T} \right] e^{- i (\omega_n t + {\bf k \cdot x})} 
,
\end{align}
where $C_n$ is a constant coefficient, and 
$\delta n_{\omega_n}, \delta T_{\omega_n}$, and $\delta {\bf u}_{\omega_n}$ are eigenfunctions obtained from the dispersion relations.

If $1 / \tau_{M*} - i \omega + i {\bf k \cdot v} = 0$,
the mode becomes continuous~\cite{book,Takamoto & Inutsuka (2010)}.
According to Eq.~(\ref{eq:mlinbolf}), 
the eigenfunction for this mode satisfies the equation
\begin{eqnarray}
0 = \delta f_{eq}
.
\end{eqnarray}
This mode represents the decay of the moments of $f$ with vanishing $\delta f_{eq}$, i.e.,
$\delta n = \delta T = 0$, $\delta {\bf u = 0}$.

Unlike the case of the nonrelativistic BGK model and Anderson-Witting model, 
the decay rate of this continuous spectrum is not constant but depends on $p^0$~\cite{cer1984}.

\subsection{\label{sec:AWlevel1}THE LINEARIZED KINETIC EQUATION AND 
DISPERSION RELATION OF THE ANDERSON-WITTING MODEL}
In this section, we derive the dispersion relation of the relativistic kinetic model of Anderson-Witting.
To obtain the dispersion relation, 
we apply an approach similar to that in the work of Cercignani and Majorana~\cite{Cercignani & Majorana(1985)}.

When there is no external field, 
the equation of the Anderson-Witting model~\cite{Anderson & Witting(1974)} is given by
\begin{align}
&\frac{D}{D s} f = p^{\mu} \partial_{\mu} f 
\label{eq:AWBGKbol}
\\ \nonumber
&= p^0 \left(\frac{\partial}{\partial t} + {\bf v} \cdot \nabla \right) f 
= - u_{\nu} p^{\nu} \frac{f-f_{eq}}{\tau} 
,
\end{align}
where $f_{eq}$ is the local equilibrium distribution function defined by Eq.~(\ref{eq:mjeq}).

As in the Marle model, 
Eq.~(\ref{eq:AWBGKbol}) is a nonlinear equation for $f(t,{\bf x, p})$ 
because of the nonlinear dependence of $f_{eq}$ on $f$ through the following matching conditions:
\begin{equation}
u_{\nu} \int \frac{d^3 p}{p^0} p^{\nu} \psi \left( f_{eq} - f \right) = 0 
\label{eq:AWmat}
,
\end{equation}
where $\psi$ is defined by Eq.~(\ref{eq:psi}).

To obtain the dispersion relation, 
we start by expanding the distribution function around a global equilibrium state $f_0({\bf p})$, 
\begin{equation}
\delta f = f - f_0, \quad \delta f_{eq} = f_{eq} - f_0. 
\end{equation}
The kinetic equation of the Anderson-Witting model Eq.~(\ref{eq:AWBGKbol}) reduces to
\begin{equation}
\left( \frac{\partial}{\partial t} + {\bf v} \cdot \nabla \right) \delta f  
= - \frac{\delta f - \delta f_{eq}}{\tau}
,
\label{eq:AWlinbol}
\end{equation}
in a linear approximation.

We assume a solution of the form
\begin{equation}
\delta f = \delta \tilde{f} e^{- i k_{\mu} x^{\mu}} = \delta \tilde{f} e^{- i \omega(t-t_0)+i {\bf k \cdot x}}
.
\label{eq:AWome}
\end{equation}

Then, Eq.~(\ref{eq:AWlinbol}) reduces to
\begin{equation}
\left( \frac{1}{\tau} -i \omega + i {\bf k \cdot v} \right) \delta f = \frac{1}{\tau} \delta f_{eq}
.
\end{equation}

We consider an equilibrium background state, 
in which the fluid is at rest, so that $u^{\mu} = (1, {\bf 0})$ and 
$\delta u^{\mu} = (0, \delta {\bf u})$ due to the relation $u^{\mu} \delta u_{\mu} = 0$.
Then, $\delta f_{eq}$ is
\begin{align}
\delta f_{eq} &= f_0 \left[ \frac{\delta n}{n} + \left(-1 + \frac{p^0}{T} + \frac{K_2'}{K_2} \zeta \right) \frac{\delta T}{T}
 - \frac{{\bf p} \cdot \delta {\bf u}}{T} \right] 
,
\\
f_0 &= \frac{n}{4 \pi m^2 T K_2(\zeta)} \exp \left[- \frac{p^0}{T} \right]
,
\end{align}
where $\delta {\bf u}$ is the space component of the Landau-Lifshitz velocity, as explained in Sec.~\ref{sec:level31}.

Using the matching conditions, we can rewrite $\delta \rho, \delta {\bf u}$, and $\delta T$ as integrals of $\delta f$:
\begin{align}
\delta n(t,{\bf x}) &= \int \frac{d^3 p}{p^0} p^0 \delta f
, \\
\delta {\bf u}(t,{\bf x}) &= - \frac{1}{n} \int \frac{d^3 p}{p^0} {\bf p} \delta f
, \\
\delta T(t,{\bf x}) &= \int \frac{d^3 p}{p^0} p^0
\\ \nonumber
&\times \frac{- 1 + K_2' \zeta / K_2 + p^0 / T}{\left(1 - K_1 \zeta / K_2 \right)
\left(3 + \zeta^2 + K_1 \zeta / K_2 \right)} \delta f
.
\end{align}

Then, Eq.~(\ref{eq:AWlinbol}) becomes
\begin{align}
&\left( \frac{1}{\tau} -i \omega + i {\bf k \cdot v} \right) \delta f({\bf p}) 
\\
\nonumber
&= \int \frac{d^3 p'}{p'^0} \frac{f_0({\bf p})}{\tau} 
\left[\frac{p'^0}{n} - \frac{{\bf p} \cdot {\bf p'}}{T}
\right.
\\ \nonumber
&+ \left. \frac{T}{n} \frac{p'^0}{\left(1-K_1 \zeta / K_2\right) \left(3 + \zeta^2 + K_1 \zeta / K_2 \right)} 
\right.
\\ \nonumber
&\times
\left.
\left(-1 + \frac{p^0}{T} + \frac{K_2'}{K_2} \zeta \right) 
 \left( - 1 + \frac{p'^0}{T} + \frac{K_2'}{K_2} \zeta \right) \right] \delta f({\bf p'})
.
\end{align}

In the following, 
we take $\tau$ as a unit of time:
\begin{equation}
\omega \tau \rightarrow \omega, \quad \tau k \rightarrow k.
\end{equation}

Finally, the linearized equation of the BGK model of Anderson-Witting is
\begin{align}
\delta f({\bf p}) &= \int \frac{d^3 p'}{p'^0} K({\bf p,p'}) \delta f({\bf p'}) 
\label{eq:AWK}
,
\\
K({\bf p,p'}) &\equiv \frac{f_0({\bf p})}{1 - i \omega + i {\bf k} \cdot \frac{{\bf p}}{p^0} }
\left[\frac{p'^0}{n} - \frac{{\bf p} \cdot {\bf p'}}{T}
\right.
\\ \nonumber
&+ \left. \frac{T}{n} \frac{p'^0}{\left(1-K_1 \zeta / K_2\right) \left(3 + \zeta^2 + K_1 \zeta / K_2 \right)} 
\right.
\\ \nonumber
&\times \left.
\left(-1 + z + \frac{K_2'}{K_2} \zeta \right) 
 \left( - 1 + z' + \frac{K_2'}{K_2} \zeta \right) \right] \delta f({\bf p'})
,
\end{align}
where $z = p^0 / T$. 
This equation make sense only when 
$1 - i \omega + i {\bf k} \cdot \frac{{\bf p}}{p^0} \neq 0$
.
When
$1 - i \omega + i {\bf k} \cdot \frac{{\bf p}}{p^0} = 0$
,
the mode becomes continuous, 
as in the Marle model.

Eq.~(\ref{eq:AWK}) is a homogeneous Fredholm integral equation of the second kind.
In particular, the kernel function $K({\bf p, p'})$ can be separated with respect to the variables ${\bf p}$
and ${\bf p'}$, and 
this equation can be solved according to a general procedure.

First, we multiply Eq.~(\ref{eq:AWK}) by $p^0$ and 
integrate with respect to ${\bf p}$.
Then, we multiply by $K_2 k$ 
and the equation reduces to
\begin{equation}
I_{11} \frac{\delta n}{n} + I_{12} {\bf k} \cdot \delta {\bf u} + I_{13} \frac{\delta T}{T} = 0
,
\label{eq:AWcont} 
\end{equation}
where
\begin{align}
I_{11} &= \zeta Q(2) - K_2 k
,
\\
I_{12} &= - \frac{i}{k} \left(3 K_2 + \zeta K_1 - b \zeta^2 Q(3) \right) 
,
\\
I_{13} &= \left(- 3 - \frac{K_1 \zeta}{K_2} \right) \zeta Q(2) + \zeta^2 Q(3)
,
\end{align}
and $Q(n)$ and $b$ are defined as follows:
\begin{align}
Q(n) &= \int^{\infty}_{1} dy \; e^{-\zeta y} y^n \arctan \frac{\sqrt{y^2 - 1}}{b \:y}
,
 \\
b &= \frac{1 - i \omega }{k}
.
\end{align}

Next, we multiply Eq.~(\ref{eq:AWK}) by $p^0 \; {\bf k \cdot p}$ and 
integrate with respect to ${\bf p}$.
Then, we multiply by $K_2$, 
and the equation reduces to
\begin{equation}
I_{21} \frac{\delta n}{n} + I_{22} {\bf k} \cdot \delta {\bf u} + I_{23} \frac{\delta T}{T} = 0 
,
\label{eq:AWmom} 
\end{equation}
where
\begin{align}
I_{21} &= 3 K_2 + \zeta K_1 - b \zeta^2 Q(3)
,
\\
I_{22} &= - i \left[\zeta K_3 - \frac{b}{k} \left\{(12 + \zeta^2) K_2 
\right. \right.
\\ \nonumber
&+ \left. \left. 3 \zeta K_1 - b \zeta^3 Q(4) \right\} \right] 
,
\\
I_{23} &= - K_1 \zeta \left(3 + \frac{K_1 \zeta}{K_2} \right) + 
(3 + \zeta^2) K_2 
\\ \nonumber
&- b \zeta^2 \left(- 3 Q(3) - \frac{K_1}{K_2} \zeta Q(3) + \zeta Q(4) \right)
,
\end{align}

Next, we multiply Eq.~(\ref{eq:AWK}) by $p^0 \; {\bf k \times p} \equiv k p^0 p^{\perp}$ and 
integrate with respect to ${\bf p}$.
Then, we multiply by $2 k K_2$, 
and the equation reduces to
\begin{equation}
I_{\perp \perp} \delta u_{\perp} = 0
,
\label{eq:AWshear}
\end{equation}
where
\begin{align}
I_{\perp \perp} &= 2 k \zeta K_3 - \zeta^3 \left\{(b^2 + 1) Q(4) - Q(2) \right\} 
\\ \nonumber
&+ b \left\{(12 + \zeta^2) K_2 + 3 \zeta K_2 \right\}
,
\end{align}

Finally, we multiply Eq.~(\ref{eq:AWK}) by $(p^0)^2$ and 
integrate with respect to ${\bf p}$.
Then we multiply by $K_2 k$, 
and the equation reduces to
\begin{equation}
I_{31} \frac{\delta n}{n} + I_{32} {\bf k} \cdot \delta {\bf u} + I_{33} \frac{\delta T}{T} = 0
,
\label{eq:AWen} 
\end{equation}
where
\begin{align}
I_{31} &= k (3 K_2 + \zeta K_1) - \zeta^2 Q(3)
,
\\
I_{32} &= - \frac{i}{k} \left[b \zeta^3 Q(4) - (\zeta^2 + 12) K_2 - 3 \zeta K_1 \right] 
,
\\
I_{33} &= k \left\{(3 + \zeta^2) K_2 - \zeta K_1 \left(3 + \frac{K_1 \zeta}{K_2} \right) \right\} 
\\ \nonumber
&- \zeta^2 \left(- 3 Q(3) - \frac{K_1}{K_2} Q(3) \zeta + \zeta Q(4) \right)
,
\end{align}

If the determinant of the above homogeneous system is set to zero,
we can obtain dispersion relation the same as Eqs. (\ref{eq:det1}), (\ref{eq:det2}), and (\ref{eq:det3}).

Using this dispersion relation, we can obtain $\delta f$ in the form
\begin{align}
\delta f({\bf v}) &= \sum_n \frac{C_n f_0({\bf v})}{1 - i \omega + i {\bf k} \cdot \frac{{\bf p}}{p^0}} 
\\ \nonumber
&\times
\left[ \frac{\delta n_{\omega_n}}{n} + \left(-1 + \frac{p^0}{T} + \frac{K_2'}{K_2} \zeta \right) \frac{\delta T_{\omega_n}}{T}
\right.
\\ \nonumber
&- \left. \frac{{\bf p} \cdot \delta {\bf u}_{\omega_n}}{T} \right] e^{- i (\omega_n t + {\bf k \cdot x})} 
,
\label{eq:AWgenesol}
\end{align}
where $C_n$ is a constant coefficient, and 
$\delta n_{\omega_n}, \delta T_{\omega_n}$, and $\delta {\bf u}_{\omega_n}$ are eigenfunctions obtained from the dispersion relations.

\section{\label{sec:level2}RESULTS}
\subsection{\label{sec:mlevel2}MARLE MODEL}
\begin{figure}[t]
 \centering
  \includegraphics[width=7cm,clip]{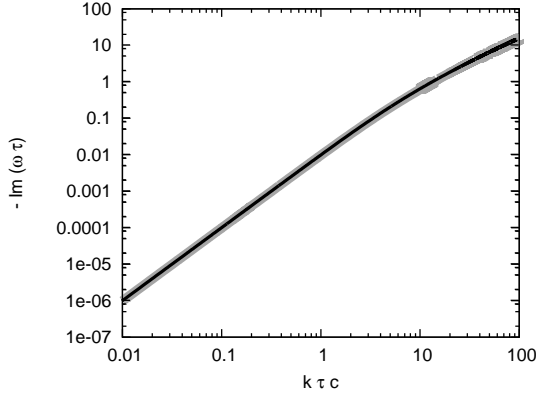}
  \caption{Decay rate of the thermal conduction mode in the nonrelativistic case: $m / T =100$.
           The black curve corresponds to the Anderson-Witting model 
           while the thick gray curve corresponds to the modified Marle model.}
  \label{fig:1}
\end{figure}

\begin{figure}[here]
 \centering
 \includegraphics[width=7cm,clip]{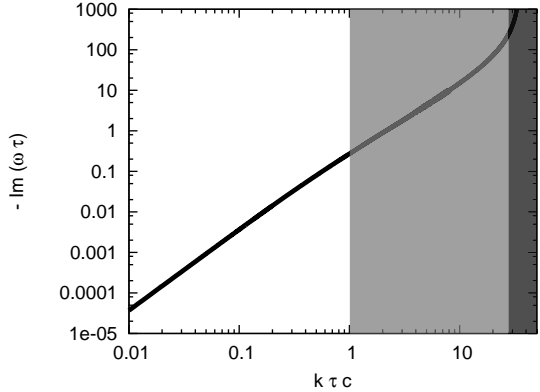}
  \caption{Decay rate of the thermal conduction mode of the modified Marle model 
          in the relativistic case: $m / T =1$.
          The gray zone represents the region in which the BGK approximation is expected to be worse;  
          the black zone represents the region in which we cannot use the BGK approximation.
          }
  \label{fig:2}
\end{figure}

\begin{figure}[t]
 \centering
 \includegraphics[width=7cm,clip]{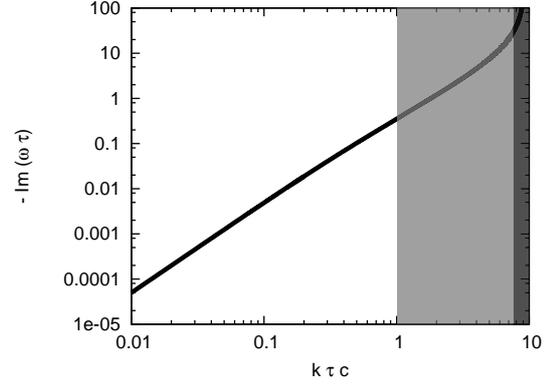}
  \caption{Decay rate of the thermal conduction mode of the modified Marle model 
          in the ultra-relativistic case: $m / T =0.01$.}
  \label{fig:3}
\end{figure}

\begin{figure}[here]
 \centering
 \includegraphics[width=7cm,clip]{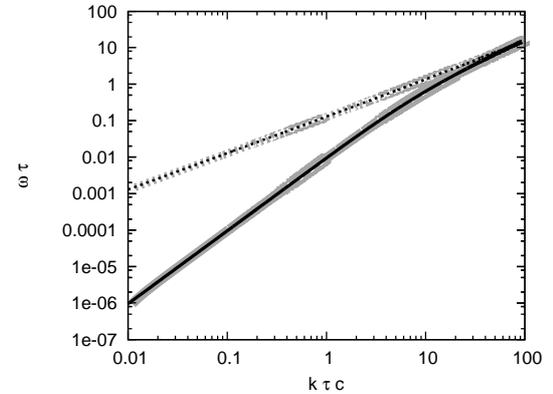}
  \caption{Dispersion relation of the sound wave mode 
           in the nonrelativistic case: $m / T = 100$.
           Solid curve represents the decay rate ($-$Im~$\omega$); 
           dotted curve represents the frequency Re~$\omega$.
           The black curve corresponds to the Anderson-Witting model 
           while the thick gray curve corresponds to the modified Marle model.}
  \label{fig:4}
\end{figure}

\begin{figure}[t]
 \centering
 \includegraphics[width=7cm,clip]{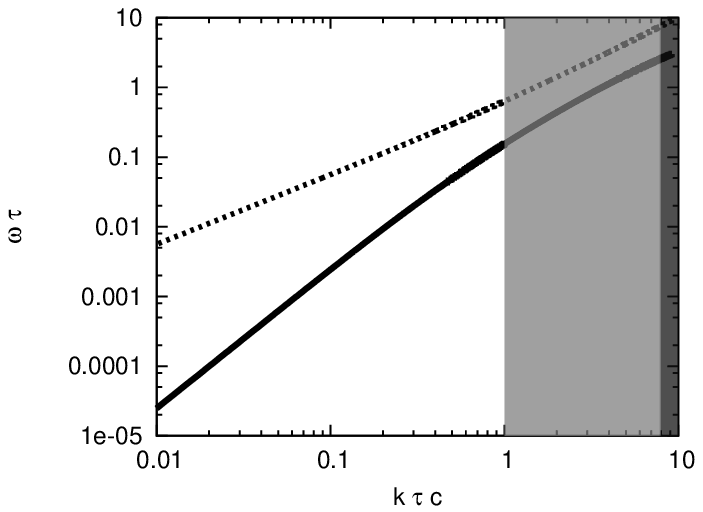}
  \caption{Dispersion relation of the sound wave mode of the modified Marle model 
           in the relativistic case: $m / T = 1$.
           Solid curve represents the decay rate ($-$Im~$\omega$); 
           dotted curve represents the frequency Re~$\omega$.}
  \label{fig:5}
\end{figure}

\begin{figure}[here]
 \centering
 \includegraphics[width=7cm,clip]{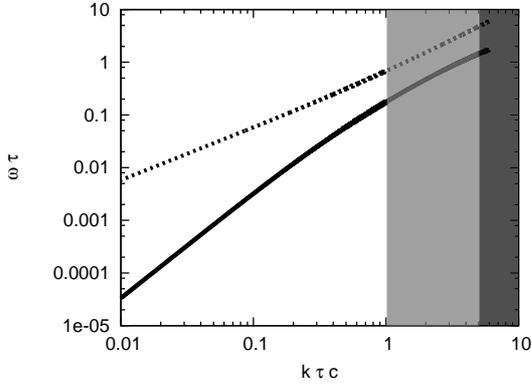}
  \caption{Dispersion relation of the sound wave mode of the modified Marle model 
           in the ultra-relativistic case: $m / T = 0.01$.
           Solid curve represents the decay rate ($-$Im~$\omega$); 
           dotted curve represents the frequency Re~$\omega$.}
  \label{fig:6}
\end{figure}

In this section, 
we show the dispersion relations of the modified Marle model obtained in the previous sections.
We solve the dispersion relations numerically; 
the results are shown below.
First, we show the thermal conduction mode in Figs.~{\ref{fig:1}}, {\ref{fig:2}}, and {\ref{fig:3}}.
At long wavelengths, 
the decay rate is proportional to $k^2$ 
and reproduces the result obtained by the first-order Chapman-Enskog approximation.
Note that the decay rate of the thermal conduction mode diverges at finite wavelengths in the relativistic cases.
This may be equivalent to the critical frequency of the thermal wave mode 
predicted in the work of Cercignani and Majorana~\cite{Cercignani & Majorana(1985)}.
We will return to this problem later.
Second, we show the sound wave mode in Figs.~{\ref{fig:4}}, {\ref{fig:5}}, and {\ref{fig:6}}.
As in the thermal conduction mode, 
at the long wavelengths the decay rate is proportional to $k^2$.
In the relativistic and ultra-relativistic cases, 
the phase velocity becomes larger than light velocity at some wavelength, 
and we stop the calculation 
because the physical collision term produces a phase speed less than light velocity~\cite{cer1983}.
As in the case of thermal conduction mode, 
this may be equivalent to the critical frequency predicted in the work of Cercignani and Majorana~\cite{Cercignani & Majorana(1985)}.
Numerically we obtain that $k_{max} \simeq 80$ when $\zeta = 5$ and $k_{max} > 100$ when $\zeta = 10$ 
where $k$ is the maximum wavelength of applicability of BGK model. 

\begin{figure}[t]
 \centering
 \includegraphics[width=7cm,clip]{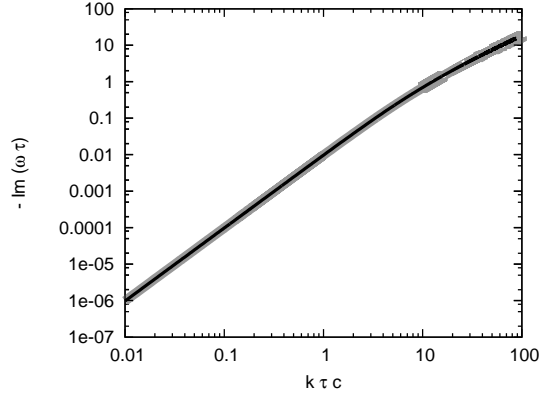}
  \caption{Decay rate of the shear flow mode  
           in the nonrelativistic case: $m / T =100$.
           The black curve corresponds to the Anderson-Witting model 
           while the thick gray curve corresponds to the modified Marle model.}
  \label{fig:7}
\end{figure}

\begin{figure}[here]
 \centering
 \includegraphics[width=7cm,clip]{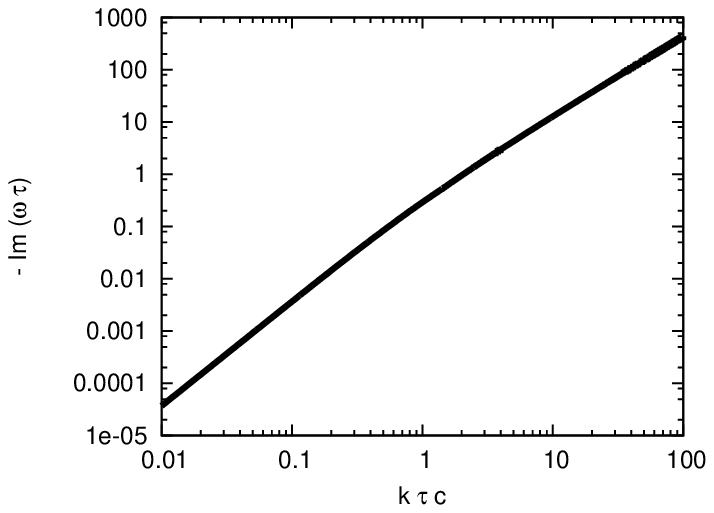}
  \caption{Decay rate of the shear flow mode of the modified Marle model 
           in the relativistic case: $m / T =1$.}
  \label{fig:8}
\end{figure}

\begin{figure}[t]
 \centering
 \includegraphics[width=7cm,clip]{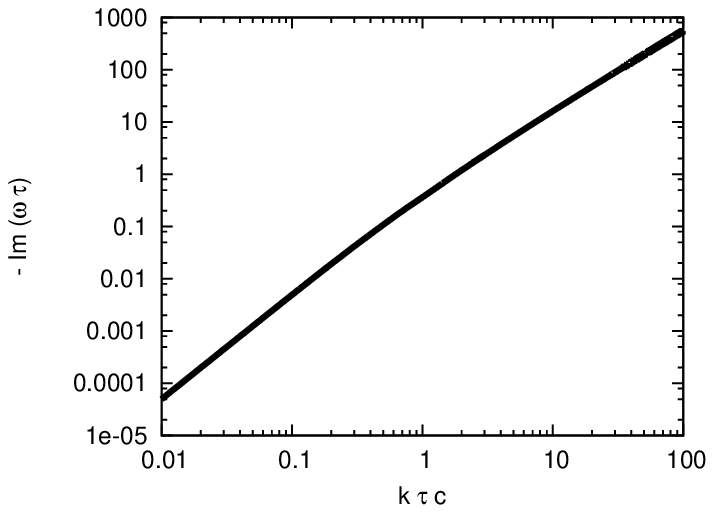}
  \caption{Decay rate of the shear flow mode of the modified Marle model 
           in the ultra-relativistic case: $m / T = 0.01$.}
  \label{fig:9}
\end{figure}

Finally, we show the shear flow mode in Figs.~{\ref{fig:7}}, {\ref{fig:8}}, and {\ref{fig:9}}.
As in the thermal conduction mode, 
at the long wavelengths the decay rate is proportional to $k^2$.
The dispersion relation for shear flow has only a decay rate, 
indicating that in rarefied gas shear flow cannot propagate.

In nonrelativistic case, 
the relevancy of the adopted equation can be checked by comparing its dispersion relations to 
experimental data of attenuation rate and phase velocity of sound wave. 
Unfortunately, corresponding experiments are very difficult in relativistic regime, 
and we cannot compare our results to experimental data. 
However, our results of nonrelativistic case ($\zeta = 100$) reproduce the dispersion relation of 
nonrelativistic BGK~\cite{Bhatnagar et al.(1954),Takamoto & Inutsuka (2010)}
that agrees with experimental data~\cite{Meyer & Sessler(1957)} even in short-wavelength regime.
For this reason, we expect that our relativistic dispersion relations should be correct even in relativistic regime, at least qualitatively.

\subsection{\label{sec:AWlevel2}ANDERSON AND WITTING MODEL}
\begin{figure}[t]
 \centering
 \includegraphics[width=7cm,clip]{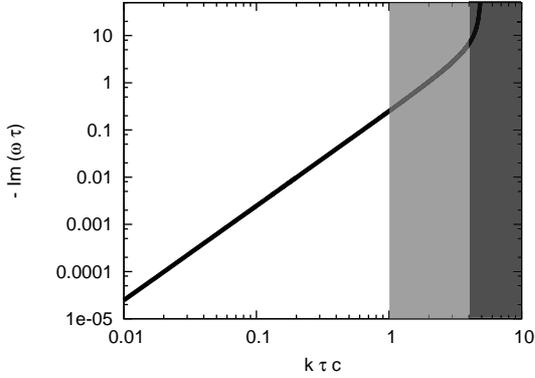}
  \caption{Decay rate of the thermal conduction mode of the Anderson-Witting model 
           in the relativistic case: $m / T =1$.}
  \label{fig:11}
\end{figure}

\begin{figure}[here]
 \centering
 \includegraphics[width=7cm,clip]{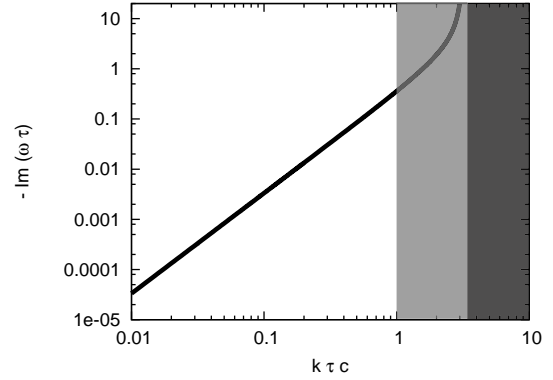}
  \caption{Decay rate of the thermal conduction mode of the Anderson-Witting model 
           in the ultra-relativistic case: $m / T =0.01$.}
  \label{fig:12}
\end{figure}

In this section, 
we show the dispersion relations of the Anderson-Witting model obtained in previous sections.
First, we show the thermal conduction mode in Figs.~{\ref{fig:11}} and {\ref{fig:12}}. 
The nonrelativistic limit $m / T = 100$ is given in Fig. {\ref{fig:1}}.
At long wavelengths, 
the decay rate is proportional to $k^2$ 
and reproduces the result obtained by the first-order Chapman-Enskog approximation.
As in the Marle model, the decay rate diverges at finite wavelengths.

\begin{figure}[here]
 \centering
 \includegraphics[width=7cm,clip]{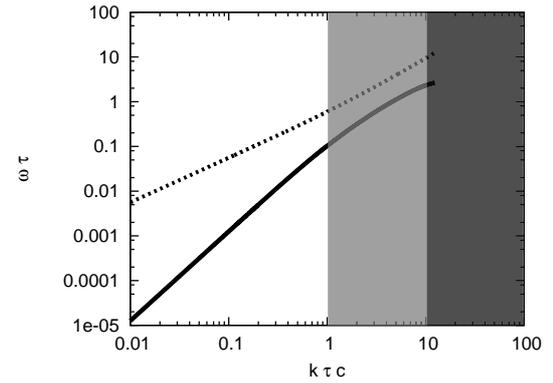}
  \caption{Dispersion relation of the sound wave mode of the Anderson-Witting model 
           in the relativistic case: $m / T = 1$.
           Solid curve represents the decay rate ($-$Im~$\omega$); 
           dotted curve represents the frequency Re~$\omega$.}
  \label{fig:14}
\end{figure}

\begin{figure}[t]
 \centering
 \includegraphics[width=7cm,clip]{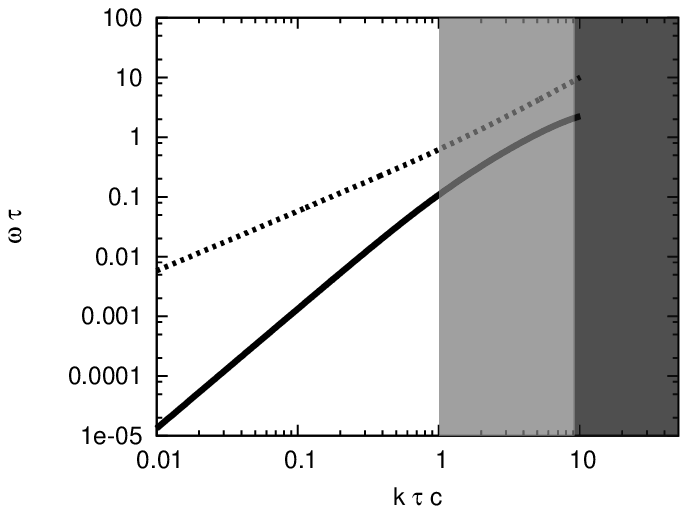}
  \caption{Dispersion relation of the sound wave mode of the Anderson-Witting model 
           in the ultra-relativistic case: $m / T = 0.01$.
           Solid curve represents the decay rate ($-$Im~$\omega$); 
           dotted curve represents the frequency Re~$\omega$.}
  \label{fig:15}
\end{figure}

Second, we show the sound wave mode in Figs.~{\ref{fig:14}} and {\ref{fig:15}}.
The nonrelativistic limit $m / T = 100$ is given in Fig. {\ref{fig:4}}.
As in the thermal conduction mode, 
at the long wavelengths the decay rate is proportional to $k^2$.
In the relativistic and ultra-relativistic case, the phase velocity becomes larger than light velocity at some wavelength, 
and we stop the calculation 
as in the Marle's model.
In this case, the wave number $k_{max}$ at which phase velocity becomes faster than light is 
 $k_{max} \simeq 64$ when $\zeta = 5$ and $k_{max} > 100$ when $\zeta = 10$.

\begin{figure}[t]
 \centering
 \includegraphics[width=7cm,clip]{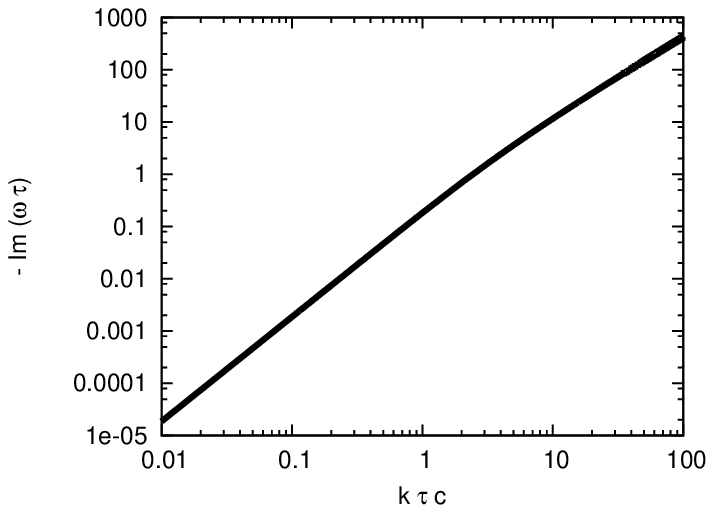}
  \caption{Decay rate of the shear flow mode of the Anderson-Witting model 
           in the relativistic case: $m / T =1$.}
  \label{fig:17}
\end{figure}

\begin{figure}[here]
 \centering
 \includegraphics[width=7cm,clip]{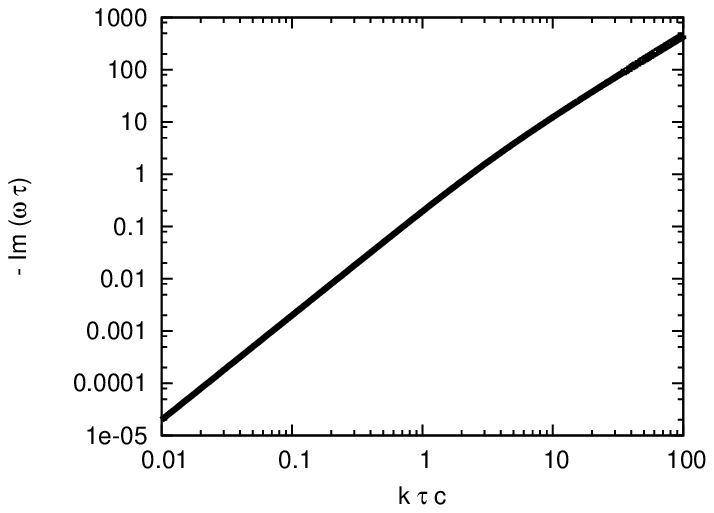}
  \caption{Decay rate of the shear flow mode of the Anderson-Witting model 
           in the ultra-relativistic case: $m / T = 0.01$.}
  \label{fig:18}
\end{figure}

Finally, we show the shear flow mode in Figs.~{\ref{fig:17}} and {\ref{fig:18}}.
The nonrelativistic limit $m / T = 100$ is given in Fig. {\ref{fig:7}}. 
As in the thermal conduction mode, 
at the long wavelengths the decay rate is proportional to $k^2$.
The dispersion relation for shear flow has only a decay rate, 
indicating that in rarefied gas shear flow cannot propagate.

\begin{figure}[t]
 \centering
 \includegraphics[width=7cm,clip]{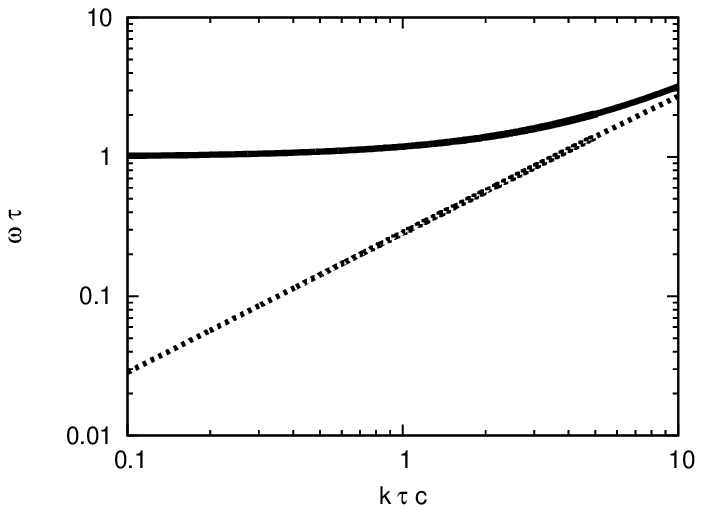}
  \caption{Dispersion relation of the transverse kinetic mode of the Anderson-Witting model 
           in the nonrelativistic case:
            $m / T = 100$.
           Solid curve represents the decay rate ($-$Im~$\omega$); 
           dotted curve represents the frequency Re~$\omega$.}
  \label{fig:19}
\end{figure}

\begin{figure}[here]
 \centering
 \includegraphics[width=7cm,clip]{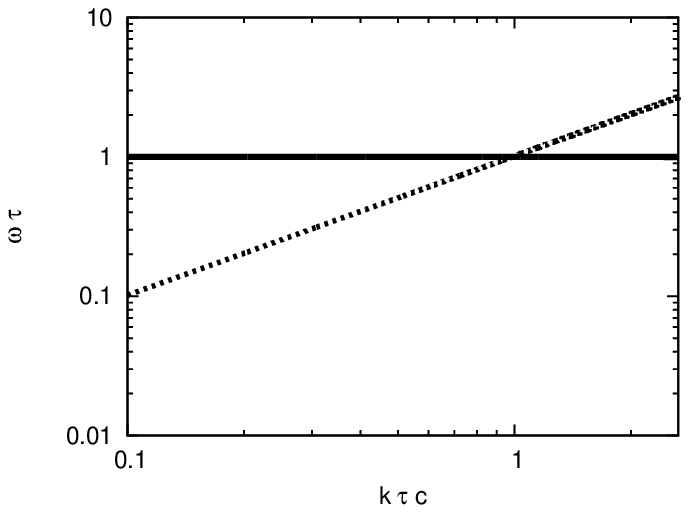}
  \caption{Dispersion relation of the transverse kinetic mode of the Anderson-Witting model 
           in the relativistic case:
            $m / T = 1$.
           Solid curve represents the decay rate ($-$Im~$\omega$); 
           dotted curve represents the frequency Re~$\omega$.}
  \label{fig:20}
\end{figure}

\begin{figure}[t]
 \centering
 \includegraphics[width=7cm,clip]{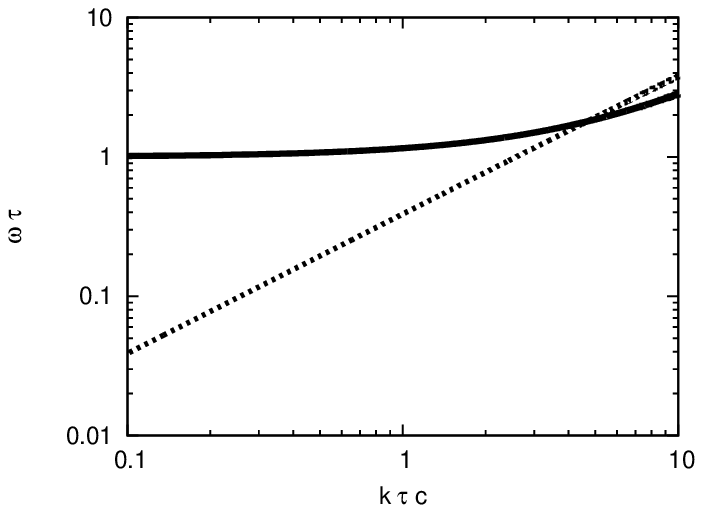}
  \caption{Dispersion relation of the longitudinal kinetic mode of the Anderson-Witting model 
           in the nonrelativistic case:
            $m / T = 100$.
           Solid curve represents the decay rate ($-$Im~$\omega$); 
           dotted curve represents the frequency Re~$\omega$.}
  \label{fig:21}
\end{figure}

\begin{figure}[here]
 \centering
 \includegraphics[width=7cm,clip]{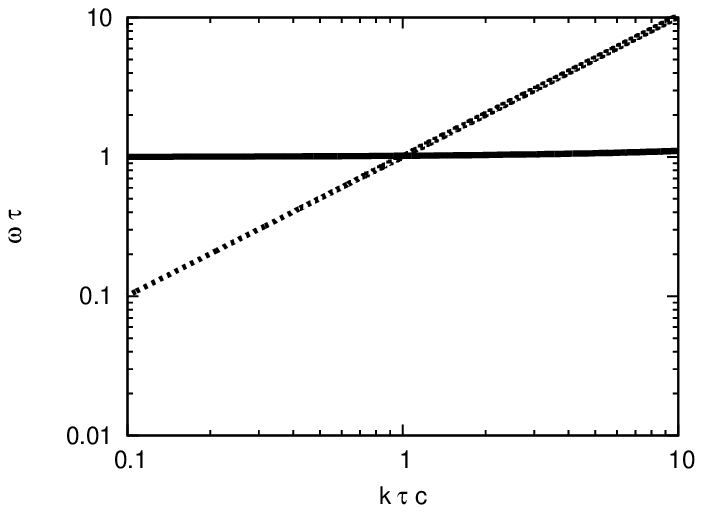}
  \caption{Dispersion relation of the longitudinal kinetic mode of the Anderson-Witting model 
           in the relativistic case:
            $m / T = 1$.
           Solid curve represents the decay rate ($-$Im~$\omega$); 
           dotted curve represents the frequency Re~$\omega$.}
  \label{fig:22}
\end{figure}

In the Anderson-Witting model, 
we find the kinetic mode. 
Figs.~\ref{fig:19} and \ref{fig:20} show the transverse kinetic mode, 
and Figs.~\ref{fig:21} and \ref{fig:22} show the longitudinal kinetic mode.
Due to the numerical difficulties, 
we cannot find the kinetic mode in the ultra-relativistic case: 
because the phase velocity of the longitudinal kinetic mode in the relativistic case is a little 
faster than that of light, 
this mode is physically incorrect.
We expect that this problem results from the accuracy of the BGK model 
and does not show that there is any longitudinal kinetic mode of in the relativistic case.

As for the Marle model, 
we cannot compare these dispersion relations and experiment; 
however, we expect that our relativistic dispersion relations is correct even in relativistic regime, 
at least qualitatively for the same reason of Marle model. 

\section{\label{sec:level3}DISCUSSION}
\subsection{\label{sec:level322}ANALYSIS IN THE SHORT WAVE LENGTH AND HIGH FREQUENCY LIMIT}
In this section, 
we analyze the dispersion relation for the short wavelength limit of the shear flow mode 
and the high frequency limit of the thermal conduction mode.
In particular, 
divergence of the decay rate of the thermal conduction mode is not observed in the analysis of the nonrelativistic BGK model,
so we have to check this divergence analytically.
For simplicity we analyze the Anderson-Witting model.

First, we analyze the dispersion relation for large wave number in the shear flow mode.
From Sec.~\ref{sec:AWlevel1}, the dispersion relation of the shear flow mode is
\begin{align}
n \delta u_{\perp} &= \int \frac{d^3 {\bf p}}{p^0} p_{\perp} \delta f 
\\
\nonumber
&= \int \frac{d^3 {\bf p}}{p^0} p_{\perp} \frac{f_{eq}}{1 - i \omega + i k v_x} 
\\
\nonumber
&\times
\left[ \frac{\delta n}{n} + \left(-1 + \frac{p^0}{T} + \frac{K_2'}{K_2} \zeta \right) \frac{\delta T}{T} + {\bf p \cdot \delta u} \right]
\\
\nonumber
&+ \frac{2 \pi}{k} \int^{\infty}_{\zeta / \sqrt{1+b^2}} d z~z^2 [(1 + b^2) z^2 - \zeta^2] f_{eq} \delta u_{\perp}
\\
\nonumber
&= \int \frac{d^3 {\bf p}}{p^0} \frac{p_{\perp} f_{eq}}{1 - i \omega + i k v_x} p_{\perp} \delta u_{\perp}
\\
\nonumber
&+ \frac{2 \pi}{k} \int^{\infty}_{\zeta / \sqrt{1+b^2}} d z~z^2 [(1 + b^2) z^2 - \zeta^2] f_{eq} \delta u_{\perp}
,
\label{eq:slwl}
\\
b &= \frac{1 - i \omega}{k}
.
\end{align}
In the above equation, 
we add correction terms of analytical continuation, 
because the decay rate $-\mathrm{Im}~\omega$ is larger than $1 / \tau$ in the short wavelength limit.

In the high frequency limit, 
we use the following approximations:
\begin{align}
\frac{1}{1 - i \omega + i k v_x} &\simeq - \frac{1}{i \omega} \left( 1 + \frac{1 + i k v_x}{i \omega} \right)
,
\\
\frac{\zeta}{\sqrt{1 + b^2}} &\simeq \frac{\zeta}{b^2}
,
\\
b &\simeq - \frac{i \omega}{k}
,
\end{align}
Using the above approximations and neglecting the terms higher than the third order of $|1 / \omega|$, 
we reduce Eq.~(\ref{eq:slwl}) to
\begin{align}
&1 - \frac{2 \pi}{k} \frac{(i \omega)^2}{k^2} \alpha = 0
,
\\
&\alpha = \int^{\infty}_{\zeta/b^2} dz~z^4 f_{eq}
\end{align}
Finally, we obtain the dispersion relation of the shear flow mode in the high frequency limit, 
\begin{equation}
\omega = - i \sqrt{2 \pi \alpha} k^{3/2}
,
\end{equation}
where we take the sign representing the decaying mode.
This reproduces the results in Sec.~\ref{sec:AWlevel2}.
In the nonrelativistic regime, 
we cannot neglect the $\omega / k$ dependence in $\alpha$, 
and the behavior of the shear flow mode in the large wave number limit becomes different from $k^{3/2}$.

Next, we consider the high frequency limit of the thermal conduction mode.
Since the decay rate diverges, 
we study only the highest-order terms in $\omega$.

From Sec.~\ref{sec:AWlevel1}, 
the conservation law for particle number Eq.~(\ref{eq:AWcont}) is
\begin{align}
\frac{\delta n}{n} &= \int \frac{d^3 {\bf p}}{p^0} p^0 \frac{f_{eq}}{1 - i \omega i k v_x} 
\\ 
\nonumber
&\times
\left[ 
\frac{\delta n}{n} + \left( - 1 + \frac{p^0}{T} + \frac{K_2'}{K_2} \zeta \right) \frac{\delta T}{T} + {\bf p \cdot \delta u}
\right]
\\
\nonumber
& + \frac{\pi}{K_2 k \zeta^2} e^{-d} 
\left[ (d^2 +2 d + 2) \frac{\delta n}{n} + i b (d^3 + 3 d^2 + 6 d) \delta u_x 
\right.
\\
\nonumber
&\left.
+ \left\{
- \left( 3 + \frac{K_1}{K_2} \zeta \right) (d^2 + 2 d + 2) 
\right. \right.
\\
\nonumber
+& \left. \left. d^3 + 3 d^2 + 6 d + 6 
\right\} \frac{\delta T}{T}
\right]
,
\\
d &\equiv \frac{\zeta}{\sqrt{1 + b^2}}
.
\end{align}
In the above equation, 
we add the correction terms of analytical continuation, 
because the decay rate $-\mathrm{Im}~\omega$ is larger than $1 / \tau$ in the large wave number limit.

As in the shear flow mode,
we expand the integrand in powers of $\omega$ and neglect terms higher than second order on the right-hand side.
Then, the above equation reduces to
\begin{align}
&\left( - 1 + \frac{2 \pi}{K_2 k \zeta^2} \right) \frac{\delta n}{n} 
\\
\nonumber
&+ \frac{6 \pi i b}{K_2 k \zeta^2} \delta u_x 
- \frac{12 \pi}{ K_2 k \zeta^2} \frac{K_1}{K_2} \zeta \frac{\delta T}{T} = 0
,
\end{align}
where we consider the relativistic limit $\zeta \ll 1$, 
so that we approximate $\exp [- d] \simeq 1$.

Similarly, the conservation of energy Eq.~(\ref{eq:AWen}) reduces to 
\begin{align}
&\left( - c_v T + \frac{6 \pi}{K_2 k \zeta^2} \right) \frac{\delta n}{n} 
\\
\nonumber
&+ \frac{24 \pi i b}{K_2 k \zeta^2} \delta u_x 
\left[
- n c_v + \frac{6 \pi}{ K_2 k \zeta^2} \left( 1 - \frac{K_1}{K_2} \zeta \right)
\right] \frac{\delta T}{T} = 0
,
\end{align}

From Sec.~\ref{sec:level31}, 
the dispersion relation of the Anderson-Witting model includes the conservation of energy Eq.~(\ref{eq:eflow}), 
so we use the conservation of energy instead of the conservation of momentum Eq.~(\ref{eq:AWmom}).
\begin{equation}
- i \omega (c_v T \delta n + n c_v \delta T) + i n h k \delta u_x = 0
,
\end{equation}
From the above equations, 
we can obtain the dispersion relation in the form
\begin{align}
&\frac{i \omega^2}{k^3 K_2^3 \zeta^4}
\left[24 c_v k K_2^2 \pi \zeta^2 
\right.
\\
\nonumber
&- \left. 12 c_v \pi^2 \left\{4 K_2 + 3 K_1 \zeta (- 1 + 8 \zeta) \right\} \right] + O (\omega) = 0,
\end{align}
In the high frequency limit, 
terms of lower order than $\omega^2$ can be neglected. 
For the left-hand side to vanish, 
the coefficient of $\omega^2$ should be $0$.
Then we obtain
\begin{equation}
k = \frac{4 K_2 - 3 K_1 \zeta + 24 K_1 \zeta^2}{2 K_2^2 \zeta^2} \pi
.
\end{equation}
If we insert $\zeta = 0.01$, we obtain $k \simeq 3.141\cdots$.
This reproduces the critical wave number for thermal conduction in the Anderson-Witting model accurately.
Similarly, if we insert $\zeta = 100$, we obtain $k \sim 10^{45}$.
This indicates that in the nonrelativistic regime the Anderson-Witting model does not effectively yield the critical wave number. 
In deriving the above equation, 
we assume $\exp[- d] \sim 1$, 
so we can not reproduce the critical wave number of $\zeta = 1$ very well.
More accurate analysis reproduces the critical wave number of $\zeta = 1$ as $k \simeq 4.958$.


\subsection{\label{sec:level33}COMPARISON TO 14-MOMENT EXPANSION}
In this section, we analyze the dispersion relation of the 14-moment theory
and compare it with that of the BGK model.
We assume that the relativistic gas is at rest, 
and we consider only longitudinal waves.
In this case, the dispersion relation is given in the work of Cercignani and Kremer\cite{Cerbook,CercignaniKremer(2001)}.
The IS equation is based on 14-moment theory; 
thus, results obtained in this section can be applied to IS equation as well.

\begin{figure}[t]
 \centering
  \includegraphics[width=7cm,clip]{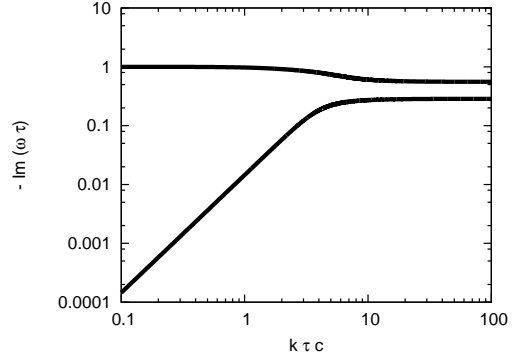}
  \caption{Decay rate of the thermal conduction and kinetic modes 
           of the 14-moment theory
           in the nonrelativistic case: $m / T =100$.}
  \label{fig:23}
\end{figure}

\begin{figure}[here]
 \centering
  \includegraphics[width=7cm,clip]{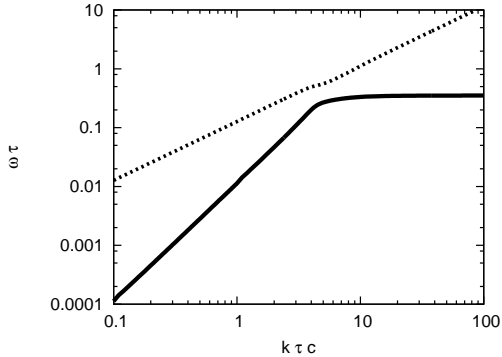}
  \caption{Dispersion relation of the sound wave mode of the 14-moment theory
           in the nonrelativistic case: $m / T = 100$.
           Solid curve represents the decay rate ($-$Im~$\omega$); 
           dotted curve represents the frequency Re~$\omega$.}
  \label{fig:24}
\end{figure}

In the nonrelativistic limiting case $\zeta \gg 1$, 
the dispersion relation reduces to
\begin{align}
&\left( \frac{k c_s}{\omega} \right)^4 \left[
\frac{567}{100} i \omega^3 - \frac{477}{100} \omega^2 - \frac{9}{10} i \omega
\right]
\\ \nonumber
&- \left( \frac{k c_s}{\omega} \right)^2 \left[
\frac{441}{50} i \omega^3 - \frac{342}{25} \omega^2 - \frac{63}{10} i \omega + 1
\right]
\\ \nonumber
&+ \frac{9}{4} i \omega^3 - \frac{21}{4} \omega^2 - 4 i \omega + 1 = 0
,
\end{align}
where
\begin{align}
c_s &= \sqrt{\frac{c_p T}{c_v h}}
,
\\
h &= m G(\zeta)
,
\\
c_v &= \zeta^2 + 5 G \zeta - G^2 \zeta^2
,
\\
c_p &= c_v + 1
,
\\
G &= \frac{K_3(\zeta)}{K_2(\zeta)}
.
\end{align}
We are interested in the Cauchy problem, 
so we solve the above equation with respect to $\omega$.
The results are illustrated in Figs.~\ref{fig:23} and \ref{fig:24} in the case of $\zeta = m / T = 100$.
Fig.~\ref{fig:23} shows the heat conduction mode and its accompanying kinetic mode. 
This figure shows that at the long wavelengths, 
the decay rate of the heat conduction mode of the 14-moment theory 
is proportional to $k^2$ and reproduces the result of the first-order Chapman-Enskog approximation. 
At short wavelengths, the decay rate of the heat conduction mode has an upper limit 
and approaches the limit asymptotically.
In addition, 14-moment theory reproduces the kinetic mode.
Fig.~\ref{fig:24} shows that the decay rate of the sound wave mode has similar features.
In comparison to the kinetic model equation and experiment~\cite{Meyer & Sessler(1957)}, 
we find that in the nonrelativistic limit, 
the behavior of the decay rate of the fluid mode of 14-moment theory
is consistent with the kinetic equation at long wavelengths 
but inconsistent at short wavelengths.
In contrast to 14-moment theory, 
the BGK approximation reproduces the result of experiment~\cite{st65} qualitatively. 
In addition, the kinetic mode obtained by 14-moment theory decrease with $k$ in contrast to kinetic modes of BGK equation. 

\begin{figure}[t]
 \centering
 \includegraphics[width=7cm,clip]{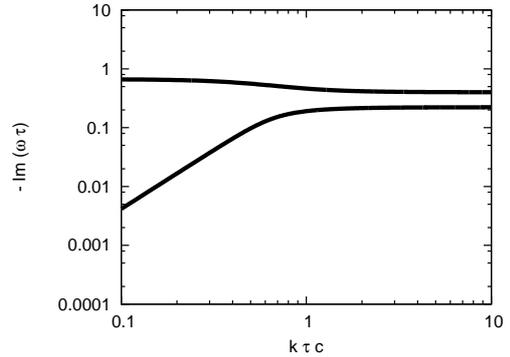}
  \caption{Decay rate of the thermal conduction and kinetic modes 
           of the 14-moment theory
           in the ultra-relativistic case: $m / T = 0.01$.}
  \label{fig:25}
\end{figure}

Next, we consider the ultra-relativistic limit $\zeta \ll 1$.
In this case, the dispersion relation reduces to
\begin{align}
&\left( \frac{k c_s}{\omega} \right)^4 \left[
\frac{225}{16} i \omega^3 - \frac{35}{4} \omega^2 - \frac{5}{4} i \omega
\right]
\\ \nonumber
&- \left( \frac{k c_s}{\omega} \right)^2 \left[
\frac{175}{8} i \omega^3 - \frac{145}{6} \omega^2 - \frac{25}{3} i \omega + 1
\right]
\\ \nonumber
&+ \frac{125}{16} i \omega^3 - \frac{145}{12} \omega^2 - \frac{73}{12} i \omega + 1 = 0
.
\end{align}
The results are illustrated in Fig.~\ref{fig:25} in the case of $\zeta = m / T = 0.01$.
Fig.~\ref{fig:25} shows the heat conduction mode and its accompanying kinetic mode.
We could not calculate the sound wave mode very accurately 
because of the numerical difficulty in the complex Newton-Raphson method, 
but the behavior of the decay rate of the sound wave mode seems to be similar to that in the nonrelativistic case.
These results indicate that at short wavelengths, 
the dispersion relation of the 14-moment theory is qualitatively different from that of the kinetic equation in the ultra-relativistic limit.

In conclusion, 
the 14-moment theory is better than the first-order Chapman-Enskog approximation 
in the sense that this theory is causal and can describe the kinetic mode.
However, the 14-moment theory cannot reproduce the result of the kinetic equation at short wavelengths 
or high frequencies, 
even in the nonrelativistic limit
in contrast to the kinetic model equations. 
This indicates that 
 the second-order dissipation terms do not reproduce kinetically correct 
 results, and keeping these second-order terms may not necessarily improve the physical description 
 of the fluid phenomena; 
 but just make the mathematical form of fluid equations 
 hyperbolic. 

%

\section{\label{sec:level4}CONCLUSION}
In this paper, we have solved the dispersion relation of the kinetic equations of the Marle 
and Anderson-Witting models with respect to $\omega$ as a function of $k$
since we are interested in the Cauchy problem.
To obtain the dispersion relation, 
an approach similar to that in the work of Cercignani and Majorana~\cite{cer1984,Cercignani & Majorana(1985)} is applied.
To obtain an acceptable dispersion relation, 
we have modified the Marle model 
since it cannot reproduce correct results in the relativistic case.
Our dispersion relation indicates that 
both kinetic model equations have a critical wavelength for the sound wave and thermal conduction modes;
for the sound wave mode, 
the phase velocity exceeds the speed of light at that wavelength~\cite{cer1983}, 
and for the thermal conduction mode, 
the decay rate diverges at that wavelength.

We have solved the dispersion relation of the 14-moment theory~\cite{Cerbook,CercignaniKremer(2001)}
with respect to $\omega$ as a function of $k$
and compared it with that of the kinetic model equations.
The results show that 
the 14-moment theory reproduces the first-order Chapman-Enskog approximation in the long wavelength region, 
but does not reproduce the result of the kinetic equation at short wavelengths 
even in the nonrelativistic limit.
This indicates that 
the second-order terms other than the time derivative of the dissipation terms are not useful 
for physical description of relativistic dissipative fluid. 

\section*{acknowledgments}
We thank Takayuki Muto and Takayuki Muranushi for fruitful discussions.
We also thank referees for advices.
This work was supported by the Grant-in-Aid for the Global COE Program 
"The Next Generation of Physics, Spun from Universality and Emergence" 
from the Ministry of Education, Culture, Sports, Science and Technology (MEXT) of Japan.


\appendix
\section{\label{sec:levela2}THE ANALYTICAL CONTINUATION}
\subsection{\label{sec:AWlevela2}THE ANDERSON AND WITTING MODEL}
In Sec.~\ref{sec:AWlevel1}, 
the integrand of the dispersion relations Eq.~(\ref{eq:AWK}) has poles.
As a result, 
when the decay rate $-$Im~$\omega$ grows to be more than $1 / \tau$, 
we must add correction terms of the analytical continuation.
In this section, 
we derive these terms.

For simplicity, 
we use the transverse shear flow mode ($\delta u_x = 0, \delta u_{\perp} \neq 0$).
The shear flow mode is given by
\begin{align}
n T \frac{K_3}{K_2} \zeta \delta u_{\perp} &= \int \frac{d^3 p}{p^0} p^0 p^{\perp} \delta f 
\label{eq:AWshear2} 
\\ \nonumber 
&= \int \frac{d^3 p}{p^0} p^0 p^{\perp} \frac{f_0}{1 - i \omega + i {\bf k} \cdot \frac{{\bf p}}{p^0} }
\\ \nonumber
&\times \left[ \frac{\delta n}{n} + \left(-1 + z + \frac{K_2'}{K_2} \zeta \right) \frac{\delta T}{T} 
\right.
\\ \nonumber
&+ \left. \left( \frac{p^x}{T} \delta u_x + \frac{p^{\perp}}{T} \delta u_{\perp} \right) \right]
\\
\nonumber
&= \int \frac{d^3 p}{p^0} p^0 p^{\perp} \frac{f_0}{1 - i \omega + i {\bf k} \cdot \frac{{\bf p}}{p^0}} p^{\perp} \delta u_{\perp}
.
\end{align}
We use cylindrical coordinates in momentum space, 
setting ${\bf k}$ as the polar axis, 
and we assume that ${\bf k}$ is parallel to the x-axis.
If we set $p_{\perp}=\sqrt{(p^y)^2 + (p^z)^2}$, 
the volume element in the momentum space $d^3 p$ reduces to
\begin{equation}
d^3 p = p_{\perp} dp_{\perp} d \phi dp^x
,
\end{equation}
where $\phi$ is the angular coordinate around the pole ${\bf k}$.

We integrate with respect to $\phi$, 
and Eq.~(\ref{eq:AWshear2}) reduces to
\begin{align}
&\frac{n T}{4 \zeta^2 K_2} \int^{\infty}_{- \infty} d p'^x \int^{\infty}_{0} d p'_{\perp} p'_{\perp} \frac{p'_{\perp} e^{-z}}
{1 - i \omega + i k p'^x/z } \times p'_{\perp} \delta u_y
\label{eq:AWshear3} \\ \nonumber
&= \frac{n T}{4 \zeta^2 K_2} \int^{\infty}_{- \infty} d p'^x \int^{\infty}_{0} d p'_{\perp} 
    \frac{z \; p'^3_{\perp} e^{-z}}{i \left\{k p'^x - i \left(1 - i \omega \right) z \right\}} \delta u_y
,
\end{align}
where
\begin{align}
p'^i &\equiv \frac{p^i}{T}
,
\\
z &\equiv \frac{p^0}{T}
.
\end{align}
This integrand has first-order poles on the real axis when Im~$\omega < -1$; 
this singularity is removed to the upper half of the complex $p^x$ plane when Im~$\omega > -1$ 
due to the collision term in the Anderson-Witting model Eq.~(\ref{eq:AWBGKbol}).
As in the Landau method, 
we consider this problem as the Cauchy problem.
Therefore, we analytically continue the integrand 
and distort the momentum contour below $i z (1 - i \omega ) / k$ in the complex $p^x$ plane.

We have to obtain the residue of the integrand of Eq.~(\ref{eq:AWshear3}).
If the integrand is $f_1(z)/f_2(z)$ and the pole is at $z_0$, 
we can use the formula
\begin{equation}
a_{-1} = \frac{f_1(z_0)}{f'_2(z_0)}, \qquad f'_2(z_0) \ne 0
,
\label{eq:formula}
\end{equation}
where $a_{-1}$ is the residue.

In Eq.~(\ref{eq:AWshear3}), $f'_2$ is given by
\begin{equation}
\frac{d}{d p'^x} \left[p'^x - i b z \right] = 1 - i b \frac{p'^x}{z}, \qquad b = \frac{1 - i \omega}{k}
.
\end{equation}
Since we consider poles, 
the denominator $f_2(z_0)$ is $0$: $p'^x - i b z = 0$.
Then, the above equation reduces to
\begin{equation}
\frac{d}{d p'^x} \left[p'^x - i b z \right] = 1 + b^2
.
\end{equation}

Using this result, we calculate the correction term of Eq.~(\ref{eq:AWshear3}): 
\begin{align}
&\frac{n T}{4 \zeta^2 K_2} \int^{\infty}_{0} d p'_{\perp} \int^{\infty}_{- \infty} d p'^x \frac{2 \pi i}{i k} \delta_D(p'^x - i b z)
 \frac{z \; p'^3_{\perp} e^{-z}}{1 + b^2} \delta u_y
\\ \nonumber
&= \left. \frac{n T}{4 \zeta^2 K_2} \frac{2 \pi}{k} \int^{\infty}_{0} d p'_{\perp} \frac{z \; p'^3_{\perp} e^{-z}}{1 + b^2} 
\right|_{p'^x - i b z = 0} \delta u_y
.
\label{eq:AWshear4}
\end{align}
Since $z^2 = (p'^x)^2 + p'^2_{\perp} + \zeta^2$, 
the equation 
\begin{equation}
p'^x - i b z = 0
\end{equation}
can be rewritten as
\begin{equation}
z = \sqrt{\frac{p'^2_{\perp} + \zeta^2}{1 + b^2}}
.
\end{equation}
Using this equation, 
we can replace the integral variable $p'_{\perp}$ with $z$.
Then, Eq.~(\ref{eq:AWshear3}) reduces to
\begin{equation}
\frac{n_0 T_0}{4 \zeta^2 K_2} \frac{2 \pi}{k} \int^{\infty}_{\frac{\zeta}{\sqrt{1 + b^2}}} d z \; z^2 
\left[(1 + b^2 ) z^2 - \zeta^2 \right] e^{-z} \delta u_y
.
\label{eq:AWshear5}
\end{equation}

We integrate Eq.~(\ref{eq:AWshear5}) and multiply by $2 K_2 k$.
Then, the equation reduces to
\begin{align}
&\frac{\pi}{2 \zeta^2} e^{- d} \left[(1 + b^2) (d^4 + 4 d^3 + 12 d^2 + 24 d + 24) 
\right.
\\
\nonumber
-& \left. d^2 (d^2 + 2 d + 2) \right] \delta u_y
,
\\
&d = \frac{\zeta}{\sqrt{1 + b^2}}
.
\end{align}
This is the correction term of the analytical continuation of Eq.~(\ref{eq:AWshear}).

In the same way, 
we can obtain the correction terms of Eqs.~(\ref{eq:AWcont}), (\ref{eq:AWmom}), and (\ref{eq:AWen}).
First, the correction term of Eq.~(\ref{eq:AWcont}) multiplied by $k K_2$ is given by
\begin{align}
&\frac{\pi}{\zeta^2} e^{- d} \left[(d^2 + 2 d + 2) \frac{\delta n}{n} 
\right.
\\
\nonumber
&+ \left. \frac{i b}{k} (d^3 + 3 d^2 + 6 d + 6) 
{\bf k} \cdot \delta {\bf u}
\right.
\\ \nonumber
&+ \left. \left\{\left(- 3 - \frac{K_1}{K_2} \zeta \right) (d^2 + 2 d + 2) + d^3 + 3 d^2 + 6 d + 6 \right\} \frac{\delta T}{T}
 \right]
.
\end{align}

Second, the correction term of Eq.~(\ref{eq:AWmom}) multiplied by $K_2$ is given by
\begin{align}
&\frac{\pi}{\zeta^2} b e^{- d} \left[(d^3 + 3 d^2 + 6 d + 6) \frac{\delta n}{n} 
\right.
\\
\nonumber
&+ \left. \frac{i b}{k} (d^4 + 4 d^3 + 12 d^2 + 24 d + 24) {\bf k} \cdot \delta {\bf u}
\right.
\\ \nonumber
&+ \left. \left\{\left(- 3 - \frac{K_1}{K_2} \zeta \right) (d^3 + 3 d^2 + 6 d + 6) 
\right. \right.
\\
\nonumber
&+ \left. \left. d^4 + 4 d^3 + 12 d^2 + 24 d + 24
\right\} \frac{\delta T}{T}
 \right]
.
\end{align}

Finally, the correction term of Eq.~(\ref{eq:AWen}) multiplied by $k K_2$ is given by
\begin{align}
&\frac{\pi}{\zeta^2} e^{- d} \left[(d^3 + 3 d^2 + 6 d + 6) \frac{\delta n}{n} 
\right.
\\
\nonumber
&+ \left. \frac{i b}{k} (d^4 + 4 d^3 + 12 d^2 + 24 d + 24) {\bf k} \cdot \delta {\bf u}
\right.
\\ \nonumber
&+ \left. \left\{\left(- 3 - \frac{K_1}{K_2} \zeta \right) (d^3 + 3 d^2 + 6 d + 6) 
\right. \right.
\\
\nonumber
&+ \left. \left. d^4 + 4 d^3 + 12 d^2 + 24 d + 24
\right\} \frac{\delta T}{T}
 \right]
.
\end{align}

%

\subsection{\label{sec:mlevela2}THE MODIFIED MARLE MODEL}
In Sec.~\ref{sec:mlevel1}, 
the dispersion relations of the modified Marle model Eqs.~(\ref{eq:mcont}), (\ref{eq:mmom}), 
(\ref{eq:mshear}), and (\ref{eq:men}) include correction terms of analytical continuation.
In this section, we derive these terms.

For simplicity, we use the transverse shear flow mode ($\delta_x = 0, \delta u_{\perp} \neq 0$).
The shear flow mode is given by
\begin{align}
n \delta u_{\perp} &= \int \frac{d^3 p}{p^0} p^{\perp} \delta f 
\label{eq:mshear2} 
\\ 
\nonumber 
&= \int \frac{d^3 p}{p^0} p^{\perp} \frac{f_0}{1 - \left(i \omega - i {\bf k} \cdot \frac{{\bf p}}{p^0} \right)
 \frac{K_1 z}{K_2 \zeta}}
\\ \nonumber
&\times \left[ \frac{\delta n}{n} + \left(-1 + z + \frac{K_2'}{K_2} \zeta \right) \frac{\delta T}{T} 
\right.
\\
\nonumber
&+ \left. \left( \frac{p^x}{T} \delta u_x + \frac{p^{\perp}}{T} \delta u_{\perp} \right) \right]
\\
\nonumber
&= \int \frac{d^3 p}{p^0} p^{\perp} \frac{f_0}{1 - \left(i \omega - i {\bf k} \cdot \frac{{\bf p}}{p^0} \right)
 \frac{K_1 z}{K_2 \zeta}} \frac{p^{\perp}}{T} \delta u_{\perp}
.
\end{align}
We use cylindrical coordinates in momentum space, setting ${\bf k}$ as the polar axis, and 
we assume that ${\bf k}$ is parallel to the x-axis.
Setting $p_{\perp}=\sqrt{(p^y)^2 + (p^z)^2}$, 
the volume element in the momentum space $d^3 p$ reduces to
\begin{equation}
d^3 p = p_{\perp} dp_{\perp} d \phi dp^x
,
\end{equation}
where $\phi$ is the angular coordinate around the pole ${\bf k}$.

We integrate with respect to $\phi$, 
and Eq.~(\ref{eq:AWshear2}) reduces to
\begin{align}
&\frac{n_0}{4 \zeta^2 K_2} \int^{\infty}_{0} d p'_{\perp} \int^{\infty}_{- \infty} d p'^x \frac{p'_{\perp}}{z} 
\frac{p'_{\perp} e^{-z}}
{1 - \left(i \omega + i k p'^x/z \right) \frac{K_1 z}{K_2 \zeta}} p'_{\perp} \delta u_y
\label{eq:mshear3} 
\\ 
\nonumber
&= \frac{n_0}{4 \zeta K_1} \int^{\infty}_{0} d p'_{\perp} \int^{\infty}_{- \infty} d p'^x \frac{p'^3_{\perp}
 e^{-z} \delta u_y}{i k z \left\{p'^x - \frac{1}{k} \left(\omega z + i \frac{K_2}{K_1} \zeta \right) \right\}}
,
\end{align}
\begin{align}
p'^i &= \frac{p^i}{T}
,
\\
z &= \frac{p^0}{T}
.
\end{align}

As in the Anderson-Witting model, 
we consider this problem as the Cauchy problem.
Therefore, we analytically continue the integrand 
and distort the momentum contour below $(\omega z + i \zeta K_2 / K_1) / k$.

We have to obtain the residue of the integrand of Eq.~(\ref{eq:mshear3}).
As in Sec.\ref{sec:AWlevela2}, 
we use Eq.~(\ref{eq:formula}).
In Eq.~(\ref{eq:mshear3}), 
$f'_2$ is given by
\begin{equation}
\frac{d}{d p'^x} \left[p'^x - \frac{1}{k} \left(\omega z + i \frac{K_2}{K_1} \zeta \right) \right] = 1 - \frac{\omega}{k} \frac{p'^x}{z}
.
\end{equation}
Since we consider the pole, the denominator $f_2(z_0)$ is $0$:
\begin{equation}
p'^x - \frac{1}{k} \left(\omega z + i \frac{K_2}{K_1} \zeta \right) = 0
.
\end{equation}
From the scalar product of the four-momentum $p^{\mu}$, we obtain
\begin{equation}
z^2 = (p'^x)^2 + p'^2_{\perp} + \zeta^2
.
\end{equation}
Using the above equations, we get
\begin{align}
p'^x =& \frac{1}{1 - \frac{\omega^2}{k^2}} \left[ i \frac{K_2 \zeta}{K_1 k} 
\right.
\\
\nonumber
&+ \left. \frac{\omega}{k} 
\sqrt{- \left(\frac{K_2 \zeta}{K_1 k} \right)^2 + (p'^2_{\perp} + \zeta^2) \left(1 - \frac{\omega^2}{k^2} \right)} \right]
,
\label{eq:manpx} \\
z =& \frac{1}{1 - \frac{\omega^2}{k^2}} \left[ i \frac{K_2 \zeta}{K_1 k} \frac{\omega}{k} 
\right.
\\
\nonumber
&+ \left. \sqrt{- \left(\frac{K_2 \zeta}{K_1 k} \right)^2 + (p'^2_{\perp} + \zeta^2) \left(1 - \frac{\omega^2}{k^2} \right)} \right]
,
\label{eq:manz}
\end{align}
where we determine the sign of $p'^x$ and $z$ to be $z \rightarrow \infty$ and Im~$p'^x < 0$ 
when $p'_{\perp} \rightarrow \infty$.

Using these equations, 
we can calculate correction terms of analytical continuation.
Considering that we have to analytically continue when $p'^x$ becomes Im~$p'^x < 0$, 
the correction terms of Eq.~(\ref{eq:mshear3}) are given by
\begin{align}
&\frac{n_0}{4 \zeta K_1} \int^{\infty}_{0} d p'_{\perp} \int^{\infty}_{- \infty} d p'^x \frac{2 \pi i}{i k z} \theta(- \mathrm{Im}~p'^x)
\\ \nonumber
&\times
\delta_D \left(p'^x - \frac{1}{k} 
\left(\omega z + i \frac{K_2}{K_1} \zeta \right) \right) \frac{p'^3_{\perp} e^{-z}}{1 - \frac{\omega}{k} \frac{p'^x}{z}} \delta u_y
\\ \nonumber
&= \left. \frac{n_0}{4 \zeta K_1} \frac{2 \pi}{k} \int^{\infty}_{0} d p'_{\perp} \theta(- \mathrm{Im}~p'^x) 
\right.
\\
\nonumber
&\times \left. \frac{p'^3_{\perp}}{z} \frac{e^{-z}}{1 - \frac{\omega}{k} \frac{p'^x}{z}} 
\right|_{p'^x - \frac{1}{k} \left(\omega z + i \frac{K_2}{K_1} \zeta \right)= 0} \delta u_y
,
\end{align}
where $\theta(x)$ is the Heaviside step function given by
\begin{eqnarray}
\theta(x) &= 0 & \mathrm{if} \quad x < 0
,
\\
\theta(x) &= 1 & \mathrm{if} \quad x > 0
.
\end{eqnarray}
Using Eq.~(\ref{eq:manz}), 
we replace the integral variables $p'_{\perp}$ with $z$.
Then, the correction terms are
\begin{equation}
\frac{n_0}{4 \zeta K_1} \frac{2 \pi}{k} \int^{\infty}_{c} d z \left[z^2 - \frac{1}{k^2} \left(z \omega + i \frac{K_2}{K_1} \zeta \right)^2 - \zeta^2
 \right] e^{-z} \delta u_y
,
\label{eq:mshear4}
\end{equation}
where $c$ is the value of $z$ evaluated when Im~$p'^x(p'_{\perp}) = 0$.
We integrate of Eq.~(\ref{eq:mshear4}) and multiply by $2 k K_1$ 
to obtain the correction terms of Eq.~(\ref{eq:mshear}).
In the same way, 
we can obtain correction terms of Eqs.~(\ref{eq:mcont}), (\ref{eq:mmom}), and (\ref{eq:men}).

%
%

\section{\label{sec:level31}RELATIONSHIP BETWEEN THE BGK MODEL AND THE MATCHING CONDITIONS}
In general, the local equilibrium distribution function lacks a physical meaning 
until it fulfills the matching conditions.
Although the matching condition of the nonrelativistic BGK model is unique, 
that of the relativistic BGK model has different forms depending on the fluid four-velocity. 
For example, the matching condition of the Marle model Eq.~(\ref{eq:mrmat}) differs from 
that of the Anderson-Witting model Eq.~(\ref{eq:AWmat}).
In this section, we consider this difference.

The kinetic equation has to fulfill the conservation laws of particle four-flow, the energy-momentum tensor. 
When we consider the kinetic equation of the BGK model, 
the conservation laws reduce to
\begin{align}
\partial_{\mu} N^{\mu} &= \int \frac{d^3 {\bf p}}{p^0} p^{\mu} \partial_{\mu} f = \int \frac{d^3 {\bf p}}{p^0} Q(f,f_{eq}) = 0
,
\\
\partial_{\mu} T^{\mu \nu} &= \int \frac{d^3 {\bf p}}{p^0} p^{\nu} p^{\mu} \partial_{\mu} f 
\\
\nonumber
&= \int \frac{d^3 {\bf p}}{p^0} p^{\nu} Q(f,f_{eq}) = 0
,
\end{align}
where $Q(f,f_{eq})$ is the collision term of the BGK model.
The collision term $Q$ depends on the local distribution function, 
so the above equations become constraints on it.
Moreover, the number of conservation laws is the same as the number of degrees of freedom of the local equilibrium distribution function.
As a result, the matching condition of the BGK model is determined by the conservation laws.

First, we consider the matching condition of the Marle model.
In this case, the above equations reduce to
\begin{align}
- \int \frac{d^3 {\bf p}}{p^0} \frac{f - f_{eq}}{\tau_{M}} &=
- \frac{1}{\tau_M} \left( \left\langle \frac{n}{e} \right\rangle - \left\langle \frac{n}{e} \right\rangle_{eq} \right) = 0
\label{eq:mm1}
,
\\
- \int \frac{d^3 {\bf p}}{p^0} p^{\nu} \frac{f - f_{eq}}{\tau_{M}} &=
- \frac{1}{\tau_M} \left( N^{\mu} - N^{\mu}_{eq} \right) = 0
\label{eq:mm2}
,
\end{align}
where $n$ is the number density of particles, 
and $e$ is the energy density per particle.
Eqs.~(\ref{eq:mm1}) and (\ref{eq:mm2}) indicate that in the Marle model, 
the conservation law of the particle four-flow Eq.~(\ref{eq:mcont}) becomes the matching condition of $n / e$, 
and the conservation laws of the energy-momentum tensor Eqs.~(\ref{eq:mmom}), (\ref{eq:mshear}), and (\ref{eq:men}) 
become the matching condition of particle number density $n$ and the Eckart velocity ${\bf u}$ .
Note that Eq.~(\ref{eq:mm1}) is not strictly the matching condition of the energy,
so that in general $e - e_{eq} \neq 0$ in the Marle model~\cite{Cerbook}.

In the nonrelativistic case, 
the conservation laws of particle four-flow are equivalent to the matching condition of the particle density $n$, 
and that of momentum flux is equivalent to the matching condition of the fluid velocity ${\bf u}$.
Thus, from these two equations we can derive a continuous equation: $- i \omega \delta n / n + i {\bf k \cdot \delta u} = 0$.
In the Marle model, however, we have to use all the equations to derive a continuous equation; 
Eq.~(\ref{eq:mcont}) $ \times i \zeta / K_1 k + $ Eq.~(\ref{eq:mmom}) $-$ Eq.~(\ref{eq:men}) $\times \omega \zeta / k$
reduces to
\begin{equation}
K_1 \zeta \omega \frac{\delta n}{n} - \zeta K_1 {\bf k \cdot \delta u} = 0
.
\end{equation}
The above equation is equivalent to the continuous equation.

Next, we consider the Anderson-Witting model.
As in the Marle model, 
we substitute $Q$ for the collision term in the Anderson-Witting model.
Then, the conservation laws reduce to
\begin{align}
- u_{\mu} \int \frac{d^3 {\bf p}}{p^0} p^{\mu} \frac{f - f_{eq}}{\tau} &=
- \frac{u_{\mu}}{\tau} \left( N^{\mu} - N^{\mu}_{eq} \right) = 0
\label{eq:AWm1}
,
\\
- u_{\mu} \int \frac{d^3 {\bf p}}{p^0} p^{\mu} p^{\nu} \frac{f - f_{eq}}{\tau} &=
- \frac{u_{\mu}}{\tau} \left( T^{\mu \nu} - T^{\mu \nu}_{eq} \right) = 0
\label{eq:AWm2}
.
\end{align}
Eqs.~(\ref{eq:AWm1}) and (\ref{eq:AWm2}) indicate that in the Anderson-Witting model, 
the conservation law of the particle four-flow Eq.~(\ref{eq:AWcont}) becomes the matching condition of particle density $n$, 
and the conservation law of the energy-momentum tensor Eqs.~(\ref{eq:AWmom}), (\ref{eq:AWshear}), and (\ref{eq:AWen}) 
becomes the matching condition of energy density $e$ and the Landau-Lifshitz velocity ${\bf u}$ .
In the Anderson-Witting model, 
the velocity is not the Eckart velocity but the Landau-Lifshitz velocity.
Therefore, we cannot derive the continuous equation from 
Eqs.~(\ref{eq:AWcont}), (\ref{eq:AWmom}), (\ref{eq:AWshear}), and (\ref{eq:AWen}) 
but we can derive the conservation law of energy $- i \omega \delta (n e) + i n h {\bf k \cdot \delta u} = 0$, 
where $h$ is the enthalpy per particle of the unperturbed gas.
To obtain this equation, 
we use only the conservation laws of energy and momentum;
Eq.~(\ref{eq:AWmom}) $\times i + $ Eq.~(\ref{eq:AWen}) $\times b$ reduces to
\begin{align}
&- i \omega (3 K_2 + \zeta K_1) \frac{\delta n}{n} 
\\
\nonumber
&+ \left[ (\zeta^2 + 3) K_2 - K_1 \zeta \left( 3 + \frac{K_1}{K_2} \zeta \right) \right] \frac{\delta T}{T} 
\\
\nonumber
&+ i \zeta K_3 {\bf k \cdot \delta u} = 0
.
\end{align}
After some calculation, 
we can check that the above equation is equivalent to the energy conservation laws:
\begin{equation}
- i \omega \delta (n e) + i n h {\bf k \cdot \delta u} = 0
.
\label{eq:eflow}
\end{equation}

\section{\label{sec:level321}ASYMPTOTIC ANALYSIS IN THE LONG WAVE LENGTH LIMIT}
To study the behavior of the roots of the dispersion relation, 
we take the long wavelength limit.
For simplicity, 
we analyze the Anderson-Witting model in this section.
First, we consider the shear flow mode.
From Sec.~\ref{sec:AWlevel1}, the dispersion relation of the shear flow is 
\begin{align}
n \delta u_{\perp} &= \int \frac{d^3 {\bf p}}{p^0} p_{\perp} \delta f 
\\
\nonumber
&= \int \frac{d^3 {\bf p}}{p^0} p_{\perp} \frac{f_{eq}}{1 - i \omega + i k v_x} 
\\
\nonumber
&\times
\left[ \frac{\delta n}{n} + \left(-1 + \frac{p^0}{T} + \frac{K_2'}{K_2} \zeta \right) \frac{\delta T}{T} + {\bf p \cdot \delta u} \right]
.
\end{align}
We expand the integrand in powers of $- i \omega + i k v_x$ and 
neglect terms higher than second order on the right-hand side.
Then, the above equation reduces to
\begin{align}
n \delta u_{\perp} 
&= \int \frac{d^3 {\bf p}}{p^0} p_{\perp} \{1 + (i \omega - i k v_x) 
\\
\nonumber
&+ (- \omega^2 - k^2 v_x^2 + 2 \omega k v_x) \} f_{eq} 
\\
\nonumber
&\times
\left[ \frac{\delta n}{n} + \left(-1 + \frac{p^0}{T} + \frac{K_2'}{K_2} \zeta \right) \frac{\delta T}{T} + {\bf p \cdot \delta u} \right]
.
\end{align}
Rewriting the above equation yields
\begin{align}
n \delta u_{\perp} 
&= \int \frac{d^3 {\bf p}}{p^0} p_{\perp} (1 + i \omega - \omega^2 - k^2 v_x^2) f_{eq} p_{\perp} \delta u_{\perp}
\\
\nonumber
&= \left[ (1 + i \omega - \omega^2) - \alpha k^2 \right] n \delta u_{\perp}
,
\\
\alpha &= \int \frac{d^3 {\bf p}}{p^0} p_{\perp}^2 v_x^2 f_{eq}
.
\end{align}
Neglecting $\omega^2$, we obtain
\begin{equation}
\omega = - i \alpha k^2
.
\end{equation}
This reproduces the dispersion relation of the shear flow mode in the long wavelength limit.

Next, we study the long wavelength limit of the thermal conduction and sound wave modes.
From Sec.~\ref{sec:AWlevel1}, the conservation of particle number Eq.(\ref{eq:AWcont}) is
\begin{align}
\frac{\delta n}{n} &= \int \frac{d^3 {\bf p}}{p^0} p^0 \frac{f_{eq}}{1 - i \omega + i k v_x} 
\\
\nonumber
&\times
\left[ 
\frac{\delta n}{n} + \left( - 1 + \frac{p^0}{T} + \frac{K_2'}{K_2} \zeta \right) \frac{\delta T}{T} + {\bf p \cdot \delta u}
\right]
.
\end{align}
As in the shear flow mode,
we expand the integrand in powers of $- i \omega + i k v_x$ and neglect terms higher than second order on the right-hand side.
Then, the above equation reduces to
\begin{align}
\frac{\delta n}{n} &= \int \frac{d^3 {\bf p}}{p^0} p^0 
\{1 + (i \omega - i k v_x) 
\\
\nonumber
&+ (- \omega^2 - k^2 v_x^2 + 2 \omega k v_x) \} f_{eq} 
\\
\nonumber
&\times
\left[ 
\frac{\delta n}{n} + \left( - 1 + \frac{p^0}{T} + \frac{K_2'}{K_2} \zeta \right) \frac{\delta T}{T} + {\bf p \cdot \delta u}
\right]
\\
\nonumber
&= (1 + i \omega - \omega^2 - \langle v_x^2 \rangle k^2 ) \frac{\delta n}{n} + (- i k T^{xx} + 2 \omega k T^{xx} ) \delta u_x 
\\
\nonumber
&+ \left[ \langle v_x^2 \rangle k^2 \left( 3 + \frac{K_1}{K_2} \zeta \right) - k^2 T^{xx} \right] \frac{\delta T}{T}
,
\\
\langle v_x^2 \rangle &\equiv \int d^3 {\bf p} v_x^2 f_{eq}
.
\end{align}
Rewriting the above equation yields
\begin{align}
&(i \omega - \omega^2 - \langle v_x^2 \rangle k^2 ) \frac{\delta n}{n} + (- i + 2 \omega ) k T^{xx} \delta u_x 
\\
\nonumber
&+ k^2 \left[ \langle v_x^2 \rangle \left( 3 + \frac{K_1}{K_2} \zeta \right) - T^{xx} \right] \frac{\delta T}{T} = 0
,
\end{align}

Similarly, the conservation of energy Eq.~(\ref{eq:AWen}) reduces to 
\begin{align}
&[ (i \omega - \omega^2 ) c_v n T - T^{xx} k^2 ] \frac{\delta n}{n} + (- i + 2 \omega ) k n h \delta u_x 
\\
\nonumber
&+ \left[ (i \omega - \omega^2 ) n T c_v 
\right.
\\
\nonumber
&+ \left. k^2 T^{xx} \left( 3 + \frac{K_1}{K_2} \zeta \right) - k^2 T^{0xx} \right] \frac{\delta T}{T} = 0
,
\end{align}
where $n$, $T$, and $h$ are the particle number density, temperature, and enthalpy of the unperturbed state, respectively, 
and $c_v$ is the heat capacity per particle.

From Sec.~\ref{sec:AWlevel1}, 
the dispersion relation of the Anderson-Witting model includes the conservation of energy Eq.~(\ref{eq:eflow}), 
so we use the conservation of energy instead of the conservation of momentum Eq.~(\ref{eq:AWmom}).
\begin{equation}
- i \omega (c_v T \delta n + n c_v \delta T ) + i n h k \delta u_x = 0
.
\end{equation}
From the above equations, 
we can obtain the dispersion relation in the form 
\begin{equation}
i A k \omega^3 + B k^3 \omega^2 + i C k^3 \omega - D k^5 = 0
,
\end{equation}
where
\begin{align}
A &= c_v n T T^{0x}
,
\\
B &= \left[2 c_v n T T^{xx} (- T^{0xx} + T^{xx} + 3 T^{xx} ) \right.
\\
\nonumber
&+ \left. T^{0x} (T^{0xx} + (- 3 + c_v n T) T^{xx} - 4 c_v n T \langle v_x^2 \rangle ) \right.
\\
\nonumber
&- \left. \frac{K_1}{K_2} (- 2 c_v n T (T^{xx})^2 
+ T^{0x} (T^{xx} + c_v n T \langle v_x^2 \rangle) ) z \right]
,
\\
C &= \left[ T^{0x} (T^{0xx} - 3 T^{xx} ) 
\right.
\\
\nonumber
&+ \left. c_v n T T^{xx} (- T^{0xx} + T^{xx} + 3 T^{xx}) \right.
\\
\nonumber
&+ \left. \frac{K_1}{K_2} T^{xx} (- T^{0x} + c_v n T T^{xx}) \zeta \right]
,
\\
D &= T^{0x} ( (T^{xx})^2 - T^{0xx} \langle v_x^2 \rangle)
.
\end{align}
We obtain three roots, denoted by $\omega_T, \omega_{S\pm}$ to second order in k: 
\begin{align}
\omega_T &= - i \frac{D}{C} k^2
,
\\
\omega_{S\pm} &= \pm \sqrt{\frac{C}{A}} k - i \frac{A~D - B~C}{2~C~A} k^2
.
\end{align}
The first solution represents the thermal conduction mode, and 
the second two represent the sound wave mode.











\end{document}